\documentclass[]{article}
\usepackage{graphicx}
\usepackage[utf8]{inputenc}
\usepackage[T1]{fontenc}
\usepackage{aeguill}
\usepackage{amssymb}
\usepackage[utf8]{inputenc}
\usepackage[T1]{fontenc}
\usepackage{aeguill}
\usepackage{amssymb}
\usepackage{graphicx}
\usepackage[a4paper,pdftex=true]{geometry}

\usepackage{natbib}
\usepackage{bbm}
\usepackage{amsmath}
\usepackage{hyperref}
\usepackage[a4paper,pdftex=true]{geometry}
\usepackage{fancyvrb}

\newtheorem{remark}{Remark}

\newcommand{\E}{E}
\newcommand{\proglang}[1]{\texttt{#1}}
\newcommand{\pkg}[1]{\texttt{#1}}
\newcommand{\VAR}{Var}

\newcommand{\Vect}[1]{\ensuremath{\mathbf{#1}}}
\newcommand{\Est}[2]{\ensuremath{\widehat{#1}\left( \Vect{#2}\right)}}
\newcommand{\Vn}{\ensuremath{\Lambda^G_{\widehat{\sigma^2},\sigma^2}}}
\newcommand{\Fn}{\ensuremath{\Lambda^G_{\widehat{r_{\sigma^2}},r_{\sigma^2}}}}
\newcommand{\EcartEst}[2][Y]{\ensuremath{\widehat{\Delta}_{\widehat{#2},#2}(\Vect{#1})}}

\newcommand{\Ecart}[2][Y]{\ensuremath{\Delta_{\widehat{#2},#2}(\Vect{#1})}}
\newcommand{\EcartG}[2][Y]{\ensuremath{\Delta^G_{\widehat{#2},#2}(\Vect{#1})}}
\newcommand{\muC}[2]{\ensuremath{\dot{\mu}_{#1}^{(#2)}}}

\title{\sc \texttt{asympTest}: an \texttt{R} package for performing parametric statistical tests and confidence intervals based on the central limit theorem}

\author{J.-F. Coeurjolly${}^{1}$, R. Drouilhet${}^1$, P. Lafaye de Micheaux${}^1$ and J.-F. Robineau${}^2$ \\
{\small ${}^1$ SAGAG Team, Laboratory Jean Kuntzmann/ CNRS, Grenoble (France), ${}^2$ CQLS, France}}

\begin{document}
\maketitle

\begin{center}
{\bf Abstract}\\
\begin{minipage}{13cm}
This paper describes an \proglang{R} package implementing large sample tests and confidence intervals (based on the central limit theorem) for various parameters. The one and two sample mean and variance 
contexts are considered.
The statistics for all the tests are expressed in the same form, which facilitates their presentation.
In the variance parameter cases, the asymptotic robustness of the classical tests depends on the departure of the data distribution from normality measured in terms of the kurtosis of the distribution.
\end{minipage}
\end{center}


\section{Introduction}

When you are interested in testing a variance parameter for a large sample  in the non Gaussian framework, it is not easy to find a test implemented in the standard statistical software. In fact, we could not find one! The only available tool is the chi-square variance test tailor-made to the Gaussian context. This test is however commonly used by practicians even if the Gaussian assumption fails. We applied it to a data set of size $n=1000$ with an empirical distribution very different from a normal distribution. The p-value of the chi-square variance test with alternative hypothesis $\mathbf{H_1}$:~$\sigma^2<1$ is 4.79\% which leads us to accept the alternative hypothesis at level $\alpha=5\%$. Is it reasonable to use this test
when we know that it cannot be used in the non Gaussian framework because of its sensitivity to departures from normality, {\it e.g.} \cite{Box53}?

From a mathematical point of view, one may wonder why no alternative test has been yet implemented since the asymptotic properties of the sample variance are well-known. Things are much easier in the sample mean study because the one sample t-test is known to be robust to departures from normality for large samples, {\it e.g.} \cite{Ozgur04}. This results from a direct application of the central limit theorem. 
The same remarks are valid when comparing the two-sample t-test (difference of means test) which is robust for large samples and the Fisher test (ratio of variances test) which is not.  \\

In the statistical framework, one may be a simple user, a tool developer, a theoretician or any combination of the above. There are natural interactions between these different communities and it is expected that their knowledge should be shared, above all for tasks that have now become very basic. It is well-known for a theoretician, see {\it e.g.}  \cite{Casella90}, that a common method for constructing a large sample test statistic may be based on an estimator that has an asymptotic normal distribution. Suppose we wish to test a hypothesis about a parameter $\theta$, and $\widehat{\theta}_n$ is some estimator of $\theta$ based on a sample of size $n$. If we can prove some form of the central limit theorem to show that as $n \to+\infty$, $(\widehat{\theta}_n-\theta )/\widehat{\sigma}_{\widehat{\theta}}\stackrel{d}{\to} \mathcal{N}(0,1)$ where $\widehat{\sigma}^2_{\widehat{\theta}}$ is a convergent (in probability) estimate of $\VAR(\widehat{\theta}_n)$, then one has the basis for an approximate test. This scheme based on the central limit theorem will be called the CLT procedure. We have already specified that we did not find any alternative to the chi-square variance test for testing a variance when the normality assumption fails. On the contrary, the problem of the robustness (to departures from normality) of tests for comparing two (or more) variances has been widely treated in the literature, see {\it e.g.} \cite{Box53}, \cite{Conover81}, \cite{Tiku86},  \cite{Pan99} and the references therein. Some alternative procedures to the Fisher test are implemented in \proglang{R}: the Bartlett test (\texttt{bartlett.test}), the Fligner test (\texttt{fligner.test}), the Levene test (\texttt{levene.test} available in the \texttt{lawstat} package), etc. However to our best knowledge, for large samples, simple alternatives based on the CLT procedure have never been proposed or implemented. \\

The main objective of this paper is to propose a unified framework, based on the CLT procedure, for large samples to test various parameters such as the mean, the variance, the difference or ratio of means or variances. This approach also allows direct derivation of asymptotic confidence intervals. Tests and confidence intervals are then implemented in our new \proglang{R} package, called \texttt{asympTest}. 
This modest contribution also solves the problem of finding a robust (to non-normality) alternative to the chi-square variance test for large samples. It also provides a very simple alternative to the Fisher test. However, note that the purpose of this paper is not to compare our tests to their competitors in terms of power. 
Finally, a first course of statistical inference usually presents mean tests in both Gaussian and asymptotical frameworks and variance tests restricted to the Gaussian case. The unified approach presented here is is very similar to the classical t-test from a mathematical point of view and gives us the opportunity to propose a more complete teaching framework with no additional difficulty. \\


The paper is organized as follows. Section~\ref{sec-math} deals with the mathematical concepts and describes our main notation. We also propose a mathematical explanation of the reason why the chi-square variance test and the Fisher test are not appropriate even when the sample size is very large. Finally, a general framework is also proposed that allows us to derive some asymptotic statistical tests for the mean, the variance and the difference (and ratio) of means or variances. In Section~\ref{sec-pkg}, the \proglang{R} package \texttt{asympTest} is presented. It notably includes the procedures described in the previous section. Finally, Sections~\ref{sec-comp},~\ref{sec-discuss} and~\ref{sec-proofs} are devoted to some discussions and the proofs of our results.

\section{Mathematical development} \label{sec-math}

\subsection{Notation}

For one-sample tests, let us denote by $\Vect{Y}=(Y_1,\ldots,Y_n)$ a sample of $n$ independent and identically distributed random variables with mean $\mu$ and variance $\sigma^2$. These parameters are classically estimated by
$$
\Est{\mu}{Y}= \overline{Y} = \frac1n \sum_{i=1}^n Y_i
\quad \mbox{ and } \quad 
\Est{\sigma^2}{Y}= \frac1{n-1}\sum_{i=1}^n \left( Y_i -\Est{\mu}{Y}\right)^2.
$$
In the two-sample context, let $\Vect{Y^{(1)}}=\left( Y_1^{(1)},\ldots,Y_{n^{(1)}}^{(1)}\right)$ and $\Vect{Y^{(2)}}=\left( Y_1^{(2)},\ldots,Y_{n^{(2)}}^{(2)}\right)$ denote two independent samples of $n^{(1)}$ and $n^{(2)}$ random variables with respective means $\mu^{(1)}$ and $\mu^{(2)}$ and variances $\sigma_{(1)}^2$ and $\sigma^2_{(2)}$. We also define the following parameters and their estimated versions by denoting $\Vect{Y}=\left(\Vect{Y}^{(1)},\Vect{Y}^{(2)}\right)$:
\begin{itemize}
\item Difference of (weighted) means: $d_{\mu}= \mu^{(1)} - \rho\times \mu^{(2)}$ ($\rho \in \mathbb{R}$) and $\Est{d_{\mu}}{\Vect{Y}}= \Est{\mu^{(1)}}{Y^{(1)}}-\rho\times \Est{\mu^{(2)}}{Y^{(2)}}$.
\item Difference of (weighted) variances: $d_{\sigma^2}= \sigma_{(1)}^2 -\rho \times \sigma_{(2)}^2$ ($\rho \in \mathbb{R}$) and $\Est{d_{\sigma^2}}{\Vect{Y}}= \Est{\sigma_{(1)}^2}{Y^{(1)}}-\rho \times \Est{\sigma_{(2)}^2}{Y^{(2)}}$.
\item Ratio of means: $r_{\mu}= \frac{\mu^{(1)}}{\mu^{(2)}}$ and $\Est{r_{\mu}}{\Vect{Y}}= \frac{\Est{\mu^{(1)}}{Y^{(1)}}}{\Est{\mu^{(2)}}{Y^{(2)}}}$.
\item Ratio of variances: $r_{\sigma^2}= \frac{\sigma_{(1)}^2}{\sigma_{(2)}^2}$ and $\Est{r_{\sigma^2}}{\Vect{Y}}= \frac{\Est{\sigma_{(1)}^2}{Y^{(1)}}}{\Est{\sigma_{(2)}^2}{Y^{(2)}}}$. 
\end{itemize}

The known parameter $\rho$ is, in our definition of $d_\mu$ and $d_{\sigma^2}$, intrinsically nonnegative. But note that there is no mathematical problem to deal with negative values of $\rho$.

In order to compare the Gaussian framework and the general one, we propose to denote by $\Vect{Y}^G$ (resp. $\Vect{Y}^{(1),G}$, $\Vect{Y}^{(2),G}$) a vector (resp. two independent  vectors) of $n$ (resp. $n^{(1)}$ and $n^{(2)}$) Gaussian random variables with mean $\mu$ (resp.  $\mu^{(1)}$ and $\mu^{(2)}$) and variance $\sigma^2$ (resp. $\sigma^2_{(1)}$ and $\sigma^2_{(2)}$). In the two-sample context, let us also denote $\Vect{Y}^G=\left( \Vect{Y}^{(1),G}, \Vect{Y}^{(2),G}\right)$.

In the sequel, we will use the notation $:=$ to define some quantity. For some random variable $Z$ and some distribution $\mathcal{L}$, $Z \leadsto \mathcal{L}$ (resp. $\stackrel{approx}{\leadsto}\mathcal{L}$) means that $Z$ follows (resp. approximately follows) the distribution $\mathcal{L}$.

\subsection{About the Chi-square and Fisher tests}

\begin{figure}
\begin{tabular}{cc}
\includegraphics[width=7cm,height=7cm]{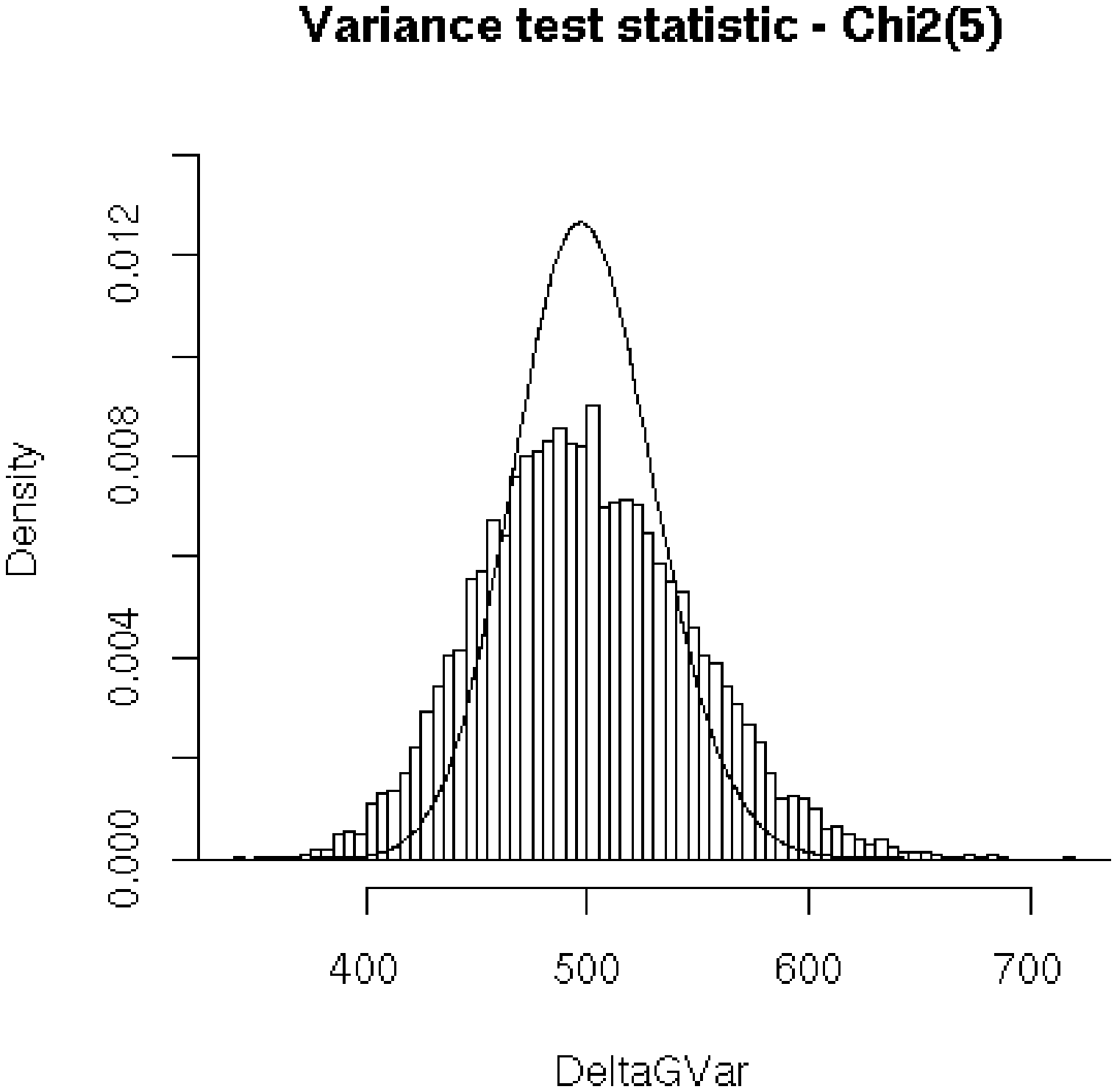}& \includegraphics[width=7cm,height=7cm]{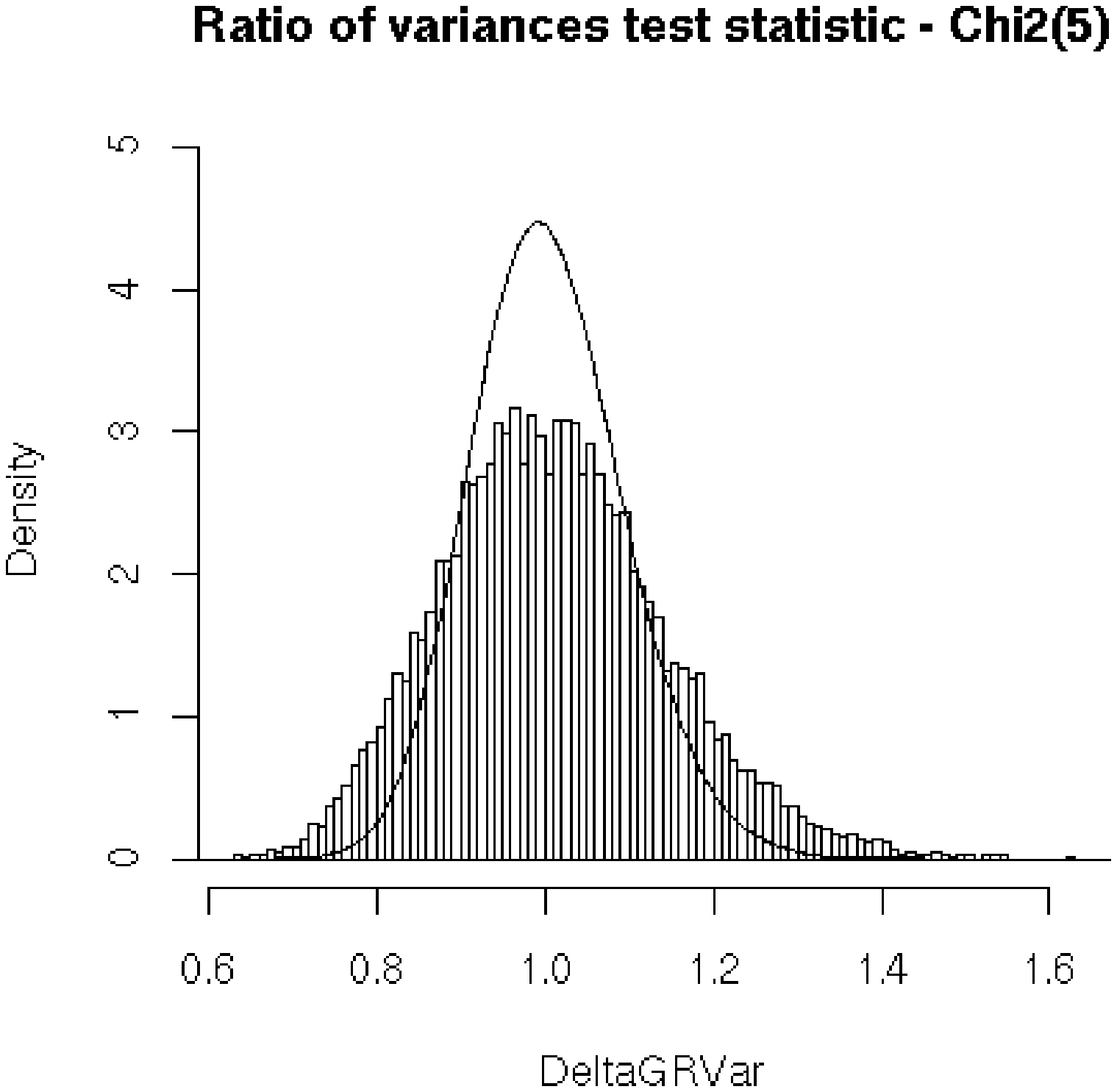} \\
\includegraphics[width=7cm,height=7cm]{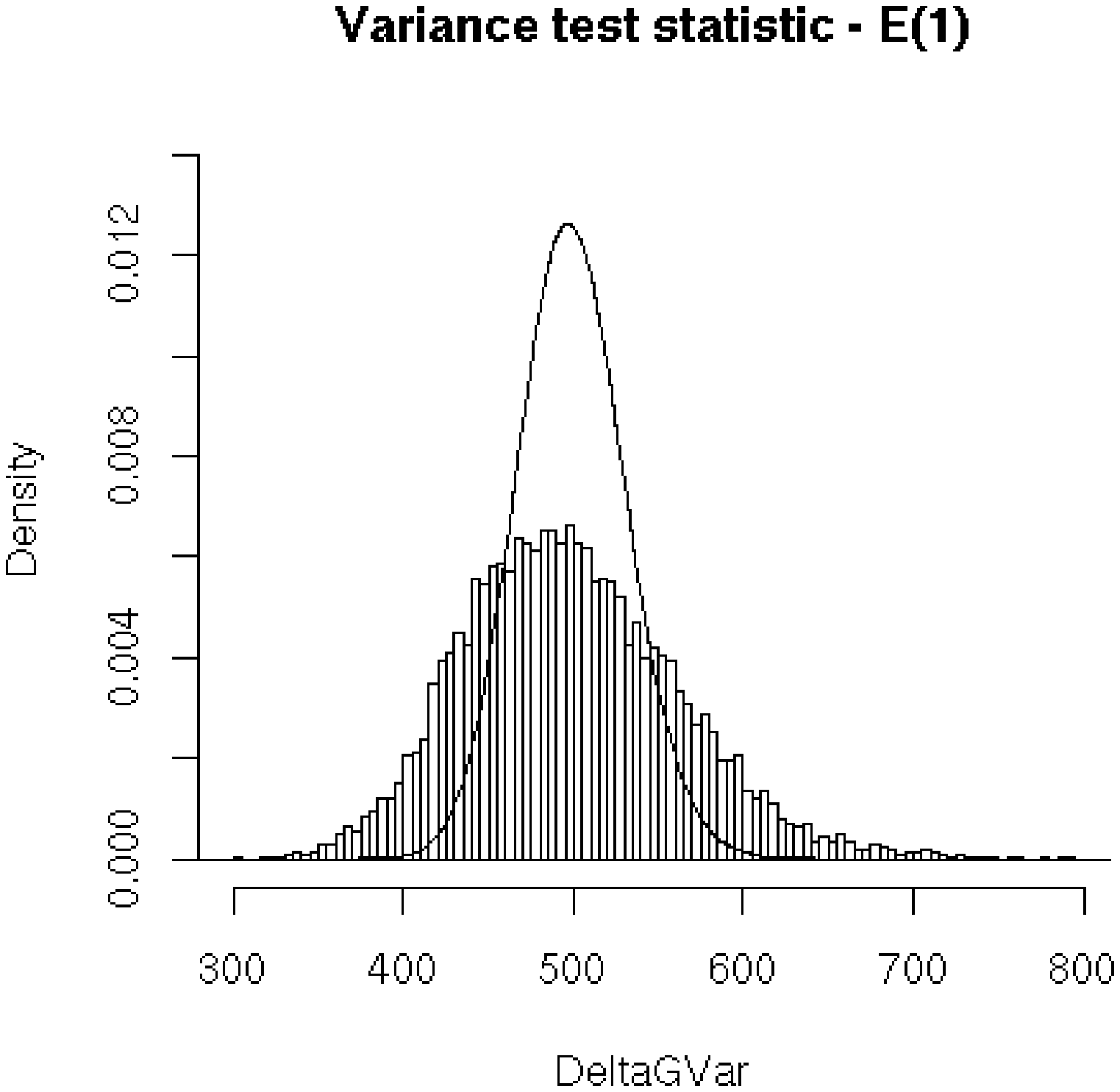}& \includegraphics[width=7cm,height=7cm]{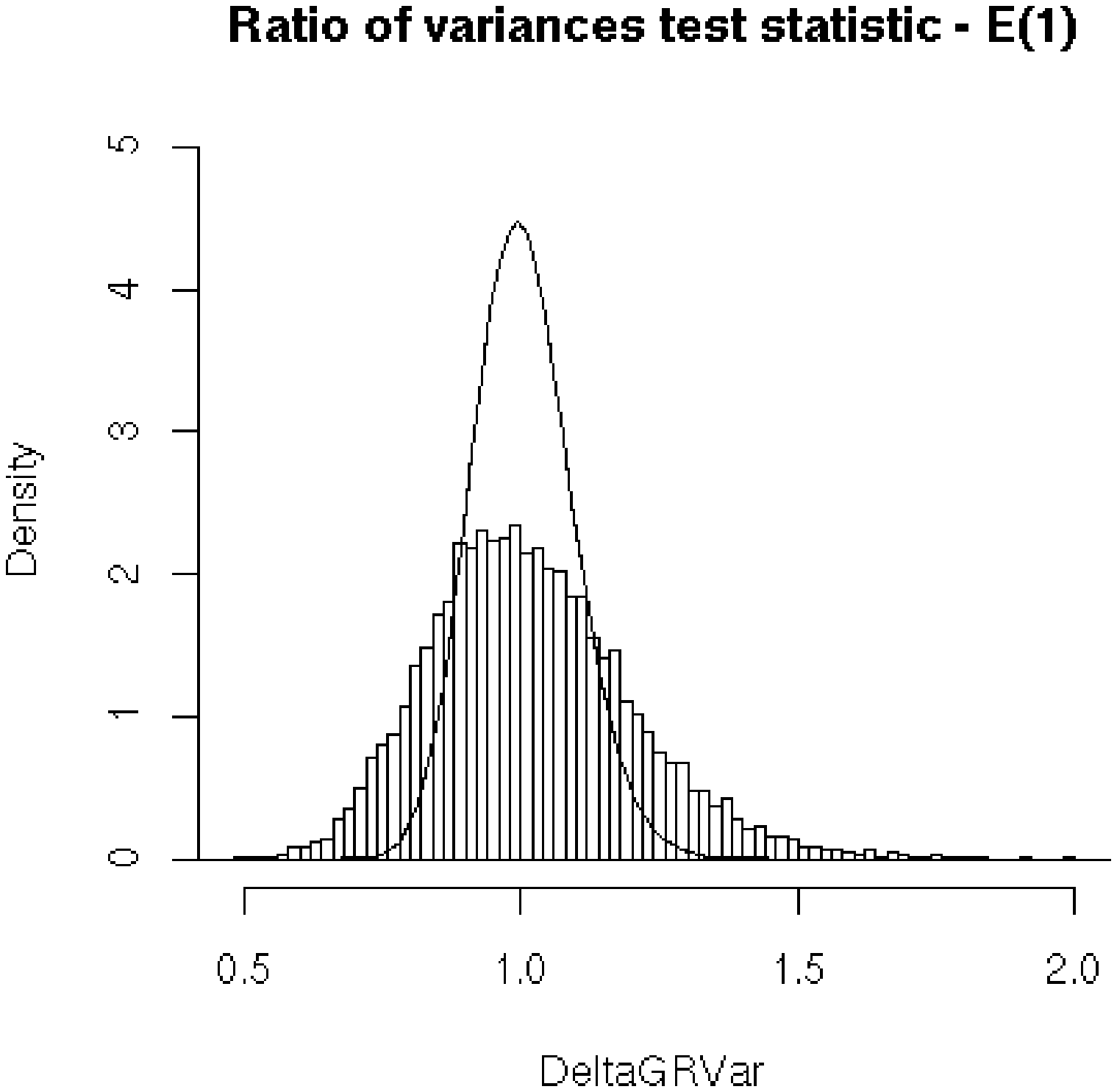} \\
\includegraphics[width=7cm,height=7cm]{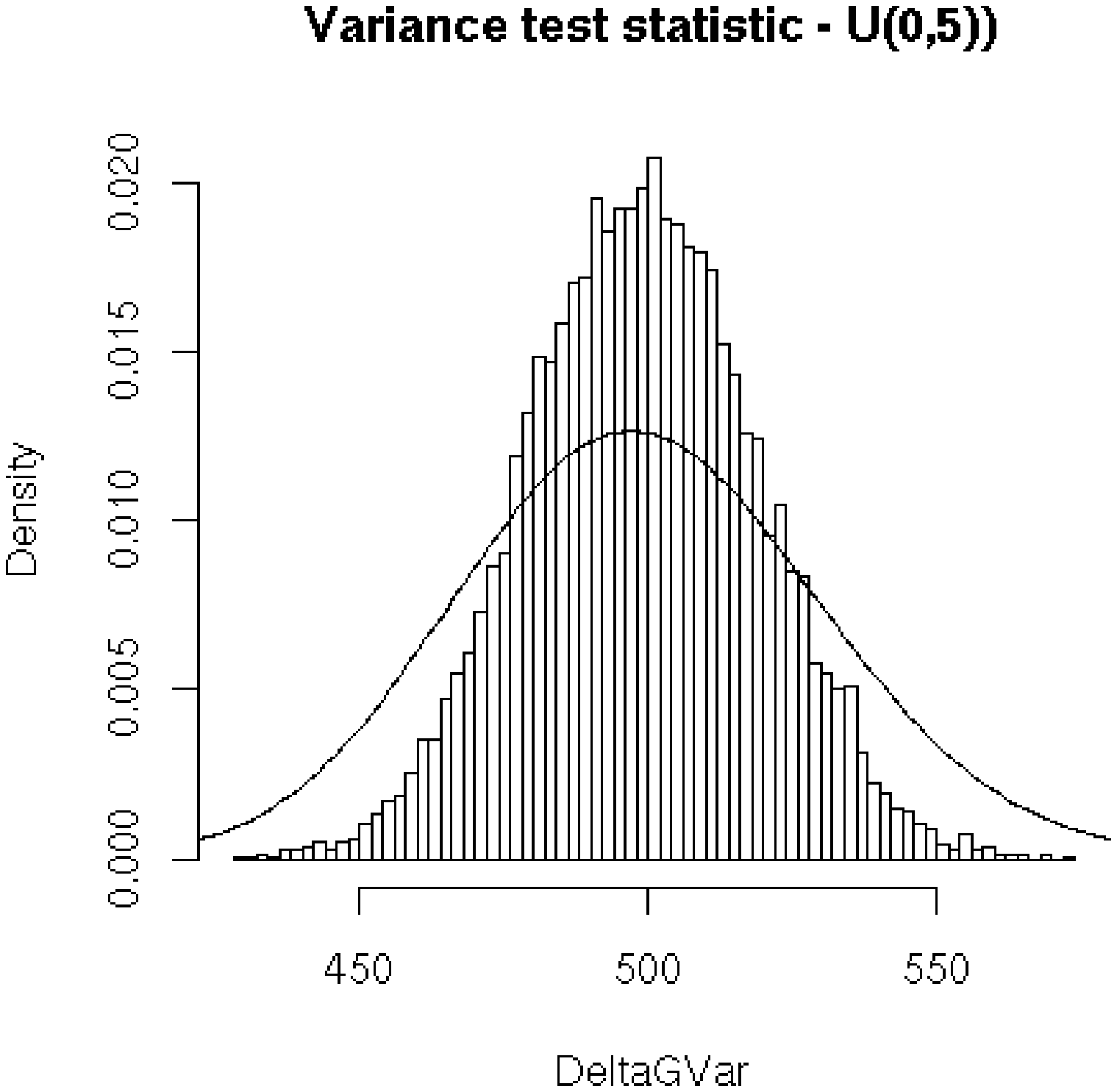}& \includegraphics[width=7cm,height=7cm]{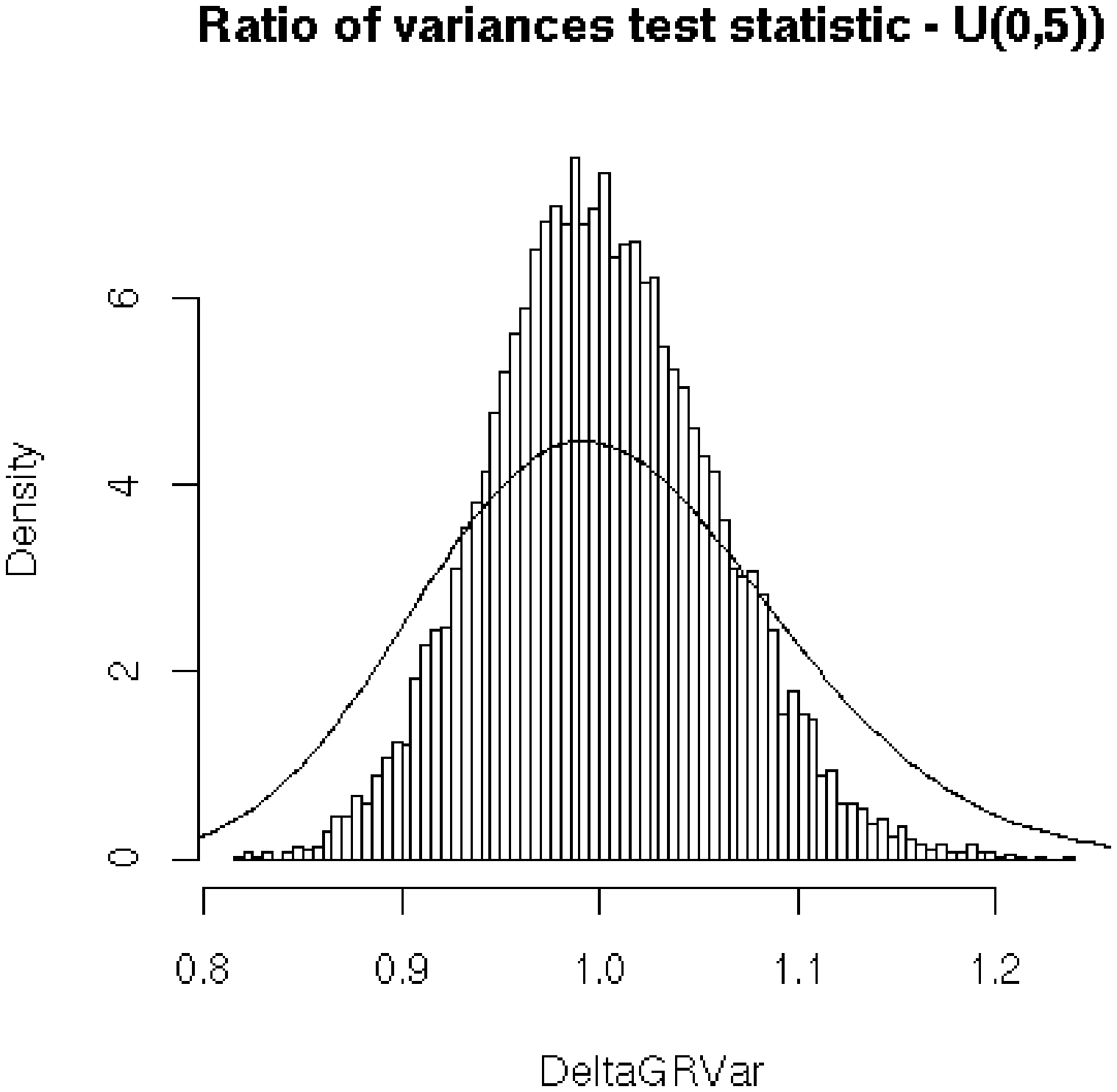} \\
\end{tabular}
\caption{Histograms of $m=10000$ replications of test statistics of variance test (left) and ratio of variance tests (right) in the Gaussian context.
The simulation has been done as follows: $n=n^{(1)}=n^{(2)}=500$, $Y\stackrel{d}{=}Y^{(1)}\stackrel{d}{=}Y^{(2)}\leadsto \chi^2(5)$ (top), $\mathcal{E}(1)$ (middle) and $\mathcal{U}([0,5])$ (bottom).} \label{fig-gauss}
\end{figure}

In this section, we concentrate on parameters $\sigma^2$ and $r_{\sigma^2}$. The classical statistics of the chi-square test of variance and the Fisher test of ratio of variances are defined by 
$$
\Vn(\Vect{Y}):=(n-1) \frac{\Est{\sigma^2}{Y}}{\sigma^2}
\qquad \mbox{ and } \qquad
\Fn(\Vect{Y}):= \frac{\Est{r_{\sigma^2}}{Y^{(1)},Y^{(2)}}}{r_{\sigma^2}}.
$$
The notation $\Vn$ expresses that the statistic is a measure of the departure of $\widehat{\sigma^2}$ from $\sigma^2$ in the Gaussian framework. When the data are Gaussian, it is well-known that 
$$
\Vn(\Vect{Y}^G) \leadsto \chi^2(n-1) 
\qquad \mbox{ and } \qquad
\Fn(\Vect{Y}^G) \leadsto  \mathcal{F}\left( n^{(1)}-1, n^{(2)}-1\right).
$$
However, as shown in Fig.~\ref{fig-gauss}, both results become untrue (even approximately) under non-normality assumption. The theoretical reason may be explained as follows. Let $n^{(1)}=n$ and assume that there exists $\alpha>0$  such that $n^{(2)}=\alpha n^{(1)}$. Assume also that $Y$ (resp. $Y^{(1)}$ and $Y^{(2)}$) has a finite kurtosis $k:=\E((Y-\mu)^4)/\VAR(Y)$ (resp. $k^{(1)}$ and $k^{(2)}$). Then, as $n \to +\infty$,
\begin{eqnarray} 
\frac{\Vn(\Vect{Y}) -(n-1)}{\sqrt{n-1}} &\stackrel{d}{\longrightarrow}& \mathcal{N}(0,k-1) \label{conv-Vn} \\
\sqrt{n}\left(\Fn(\Vect{Y})-1 \right)&\stackrel{d}{\longrightarrow}& \mathcal{N}\left(0,k^{(1)}-1 + \frac{k^{(2)}-1}\alpha\right) \label{conv-Fn}.
\end{eqnarray}
This result is a consequence of a more general one stated in Section~\ref{sec-tcl} and proved in Section~\ref{sec-proofs}. Equations~(\ref{conv-Vn}) and~(\ref{conv-Fn}) lead to the following approximations
$$
\Vn(\Vect{Y}) \stackrel{approx}{\leadsto} \mathcal{N}\left(n-1,(k-1)(n-1) \right)
\; \mbox{ and } \;
\Fn(\Vect{Y})\stackrel{approx}{\leadsto}
\mathcal{N}\left(1,\frac{k^{(1)}-1}{n^{(1)}}+ \frac{k^{(2)}-1}{n^{(2)}} \right).
$$
When data are Gaussian ans when $n$ is large, one obtains the well-known approximations 
$$
\chi^2(n-1) \simeq \mathcal{N}\left( n-1, 2(n-1) \right) 
\qquad \mbox{ and } \qquad 
\mathcal{F}\left(n^{(1)},n^{(2)}\right) \simeq \mathcal{N}\left( 1, \frac{2}{n^{(1)}}+ \frac{2}{n^{(2)}} \right)
$$
since $k=k^{(1)}=k^{(2)}=3$
We can underline that $\Vn(\Vect{Y})$ (resp. $\Fn(\Vect{Y})$) and $\Vn(\Vect{Y}^G)$ (resp. $\Fn(\Vect{Y}^G)$) differ in terms of asymptotical variances. More precisely, the gap between the two frameworks is essentially governed by the kurtosis. Indeed, as $n\to +\infty$
$$
\frac{\VAR( \Vn(\Vect{Y})) }{ \VAR( \Vn(\Vect{Y}^G))} \rightarrow \frac{k-1}{2} 
\qquad \mbox{ and } \qquad 
\frac{\VAR( \Fn(\Vect{Y}))}{\VAR( \Fn(\Vect{Y}^G))} \rightarrow \frac{k^{(1)}-1 + \frac{k^{(2)}-1}\alpha}{2\left(1+\frac1\alpha\right)}.
$$
Tab.~\ref{tab-kurt} proposes the computations of these asymptotic ratios for different distributions. This allows the reader to assess the risk of using the classical statistics $\Vn$ and $\Fn$ under the non-normality assumption even when the size of the sample is large.

\begin{table}[htbp]
\begin{tabular}{|c|c|c|c|}
\cline{2-4} 
\multicolumn{1}{c}{}&\multicolumn{3}{|c|}{$Y\stackrel{d}{=}Y^{(1)}\stackrel{d}{=}Y^{(2)} \leadsto \mathcal{L}$} \\
\hline
Test& $\quad \mathcal{L}=\chi^2(\nu)\quad $ &  $\quad \mathcal{L}=\mathcal{E}(\lambda)\quad $ & $\quad \mathcal{L}=\mathcal{U}([a,b])\quad $ \\
\hline 
&&&\\Variance test& $1+\frac6{\nu}$ & 4& $\frac25$ \\&&&\\
\hline
&&&\\Ratio of variances test & $1+\frac6{\nu}$ & 4& $\frac25$ \\
&&&\\\hline
\end{tabular}
\caption{Ratio of asymptotic variances (non gaussian/gaussian) $\frac{k-1}2$ and $\frac{ k^{(1)}-1+ \frac{k^{(2)}-1}\alpha}{2\left(1+\frac1\alpha\right)}$ in the case where $k=k^{(1)}=k^{(2)}$ and $\alpha=1$.} \label{tab-kurt}
\end{table}

\subsection{Large sample tests based on the central limit theorem}
\label{sec-tcl}

The parameters  $\sigma^2$, $d_{\mu}$ and $d_{\sigma^2}$ can be viewed as particular means. Therefore, the idea (widely used in asymptotic theory) is to design asymptotic tests and confidence intervals thanks to the central limit theorem. The variables $\Est{r_\mu}{Y}-r_\mu$ and $\Est{r_{\sigma^2}}{Y}-r_{\sigma^2}$ can also be expressed in terms of means to which a central limit theorem can be applied. In order to unify asymptotic results, we define $\theta$ as one of the parameters $\mu$, $\sigma^2$, $d_\mu$, $d_{\sigma^2}$, $r_\mu$ and $r_{\sigma^2}$. By applying a central limit theorem, the law of large numbers and Slutsky's theorem (see Section~\ref{sec-proofs}), one obtains, as $n\to +\infty$,
\begin{equation}\label{eq-tcl}
\EcartEst{\theta} :=\frac{\Est{\theta}{Y} - \theta}{\Est{\sigma_{\widehat{\theta}}}{Y}} \stackrel{d}{\longrightarrow} \mathcal{N}(0,1),
\end{equation}
where $\Est{\sigma_{\widehat{\theta}}}{Y}$ is the standard error of $\Est{\theta}{Y}$. The assumptions under which the central limit theorem can be applied and the definition of $\Est{\sigma_{\widehat{\theta}}}{Y}$ are stated in Tab.~\ref{tab-resTCL}. \proglang{R} functions have been implemented to evaluate the standard errors of the estimates of $\theta$, see Tab.~\ref{tab-se}.

\begin{table}[htbp]
\begin{center}
\begin{tabular}{|c|c|c|}
\hline
$\theta$ & Assumptions & $\Est{\sigma_{\widehat{\theta}}}{Y}$ \\
\hline
$\mu$ & $\E \left(Y_i^2 \right)<+\infty$&  $\displaystyle{\sqrt{ \frac{\Est{\sigma^2}{Y} }{n}}}$\\
\hline
$\sigma^2$& $\E \left(Y_i^4 \right)<+\infty$ & $\displaystyle{\sqrt{ \frac{\Est{\sigma^2_{\ddot{Y}}}{\ddot{Y}}}{n}}}$ \\
\hline
$d_{\mu}$ & $\E \left( \left(Y_i^{(j)}\right)^2 \right)<+\infty$, $j=1,2$&
$\displaystyle{ \sqrt{ \frac{\Est{\sigma^2_{(1)}}{Y^{(1)}}}{n^{(1)}} + \rho^2 \times \frac{\Est{\sigma^2_{(2)}}{Y^{(2)}}}{n^{(2)}}}}$\\
\hline
$d_{\sigma^2}$ &$\E \left( \left(Y_i^{(j)}\right)^4 \right)<+\infty$, $j=1,2$ & $\displaystyle{ \sqrt{ \frac{\Est{\sigma^2_{\ddot{Y}^{(1)}}}{\ddot{Y}^{(1)}}}{n^{(1)}} +\rho^2\times\frac{\Est{\sigma^2_{\ddot{Y}^{(2)}}}{\ddot{Y}^{(2)}}}{n^{(2)}}}}$\\
\hline 
$r_\mu$& $\mu^{(2)}\neq 0$, $\E \left( \left(Y_i^{(j)}\right)^2 \right)<+\infty$, $j=1,2$&
${\frac1{\left| \Est{\mu^{(2)}}{Y^{(2)}} \right|} \sqrt{\frac{\Est{\sigma_{(1)}^2}{Y^{(1)}}}{n^{(1)}}+\Est{r_{\mu}}{Y}^2\times\frac{\Est{\sigma_{(2)}^2}{Y^{(2)}}}{n^{(2)}}}}$ \\
\hline
$r_{\sigma^2}$ &$\sigma^2_{(2)}\neq 0$, $\E \left( \left(Y_i^{(j)}\right)^4 \right)<+\infty$, $j=1,2$& ${\frac1{\Est{\sigma^2_{(2)}}{Y^{(2)}}} \sqrt{\frac{\Est{\sigma^2_{\ddot{Y}^{(1)}}}{\ddot{Y}^{(1)}}}{n^{(1)}}\!+\!\Est{r_{\sigma^2}}{Y}^2   \frac{\Est{\sigma^2_{\ddot{Y}^{(2)}}}{\ddot{Y}^{(2)}}}{n^{(2)}}}}$\\
\hline
\end{tabular}
\caption{Standard errors of estimates of $\theta$. For the sake of simplicity we denote by $\ddot{Y}:=(Y-\mu)^2$, $\ddot{\Vect{Y}}:=\left( \Vect{Y} -\Est{\mu}{Y}\right)^2$, $\sigma^2_{\ddot{Y}}:=\VAR\left( \ddot{Y}\right)$ and $\Est{\sigma_{\ddot{Y}}^2}{\ddot{\Vect{Y}}}:=\frac1{n-1}\sum_{i=1}^n\left( \left(Y_i-\Est{\mu}{Y}\right)^2- \Est{\sigma^2}{Y}\right)^2 $.  }\label{tab-resTCL}
\end{center}
\end{table}

\begin{table}[htbp]
\begin{center}
\begin{tabular}{|c|c|c|}
\hline
\hspace*{1cm}$\theta$ \hspace*{1cm}& \hspace*{1cm}Dataset(s) \hspace*{1cm}& \hspace*{1cm}$\Est{\sigma_{\widehat{\theta}}}{y}$ in \proglang{R}\hspace*{1cm} \\
\hline
$\mu$ & \texttt{y}&  \texttt{seMean(y)}\\
\hline
$\sigma^2$& \texttt{y} & \texttt{seVar(y)} \\
\hline
$d_{\mu}$ & \texttt{y1}, \texttt{y2}& \texttt{seDMean(y1,y2,rho=1)} \\
\hline
$d_{\sigma^2}$ &\texttt{y1}, \texttt{y2}& \texttt{seDVar(y1,y2,rho=1)}\\
\hline
$r_\mu$&\texttt{y1}, \texttt{y2}& \texttt{seRMean(y1,y2)} \\
\hline
$r_{\sigma^2}$ &\texttt{y1}, \texttt{y2}& \texttt{seRVar(y1,y2)} \\
\hline
\end{tabular}
\end{center}
\caption{Standard errors of estimates of $\theta$ in \proglang{R}.}\label{tab-se}
\end{table}

\begin{figure}
\begin{tabular}{cc}
\includegraphics[width=7cm,height=7cm]{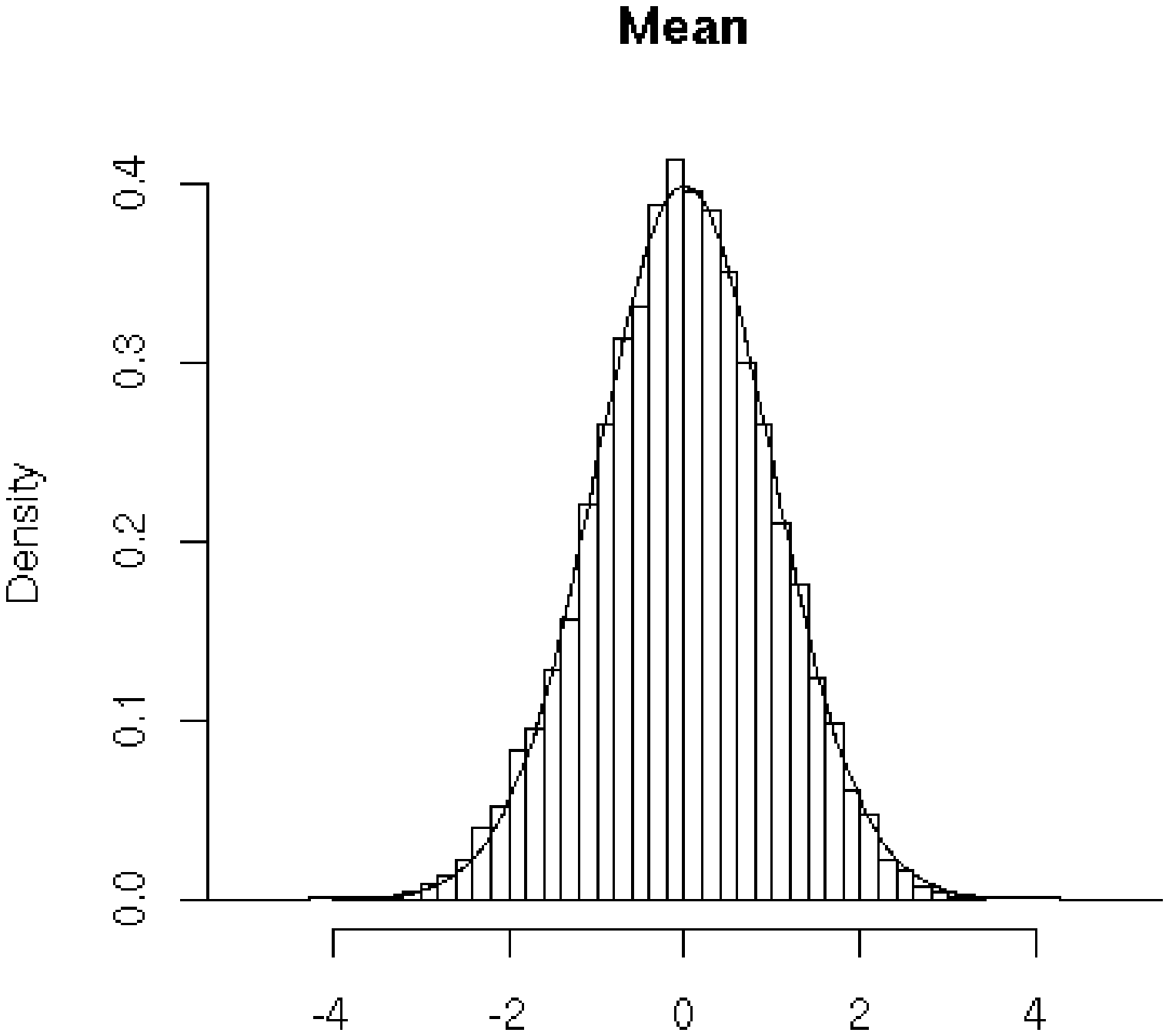}& \includegraphics[width=7cm,height=7cm]{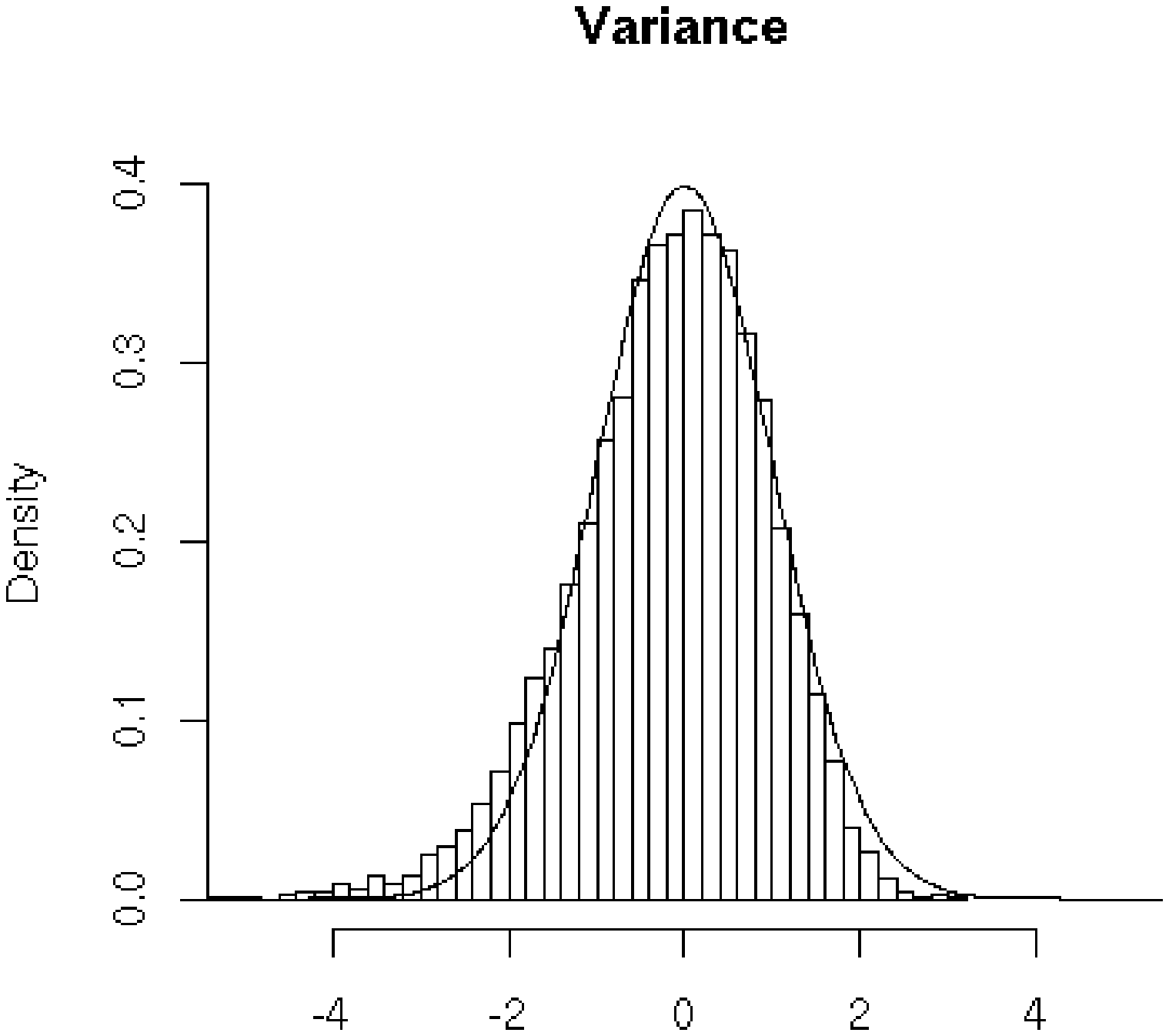} \\
\includegraphics[width=7cm,height=7cm]{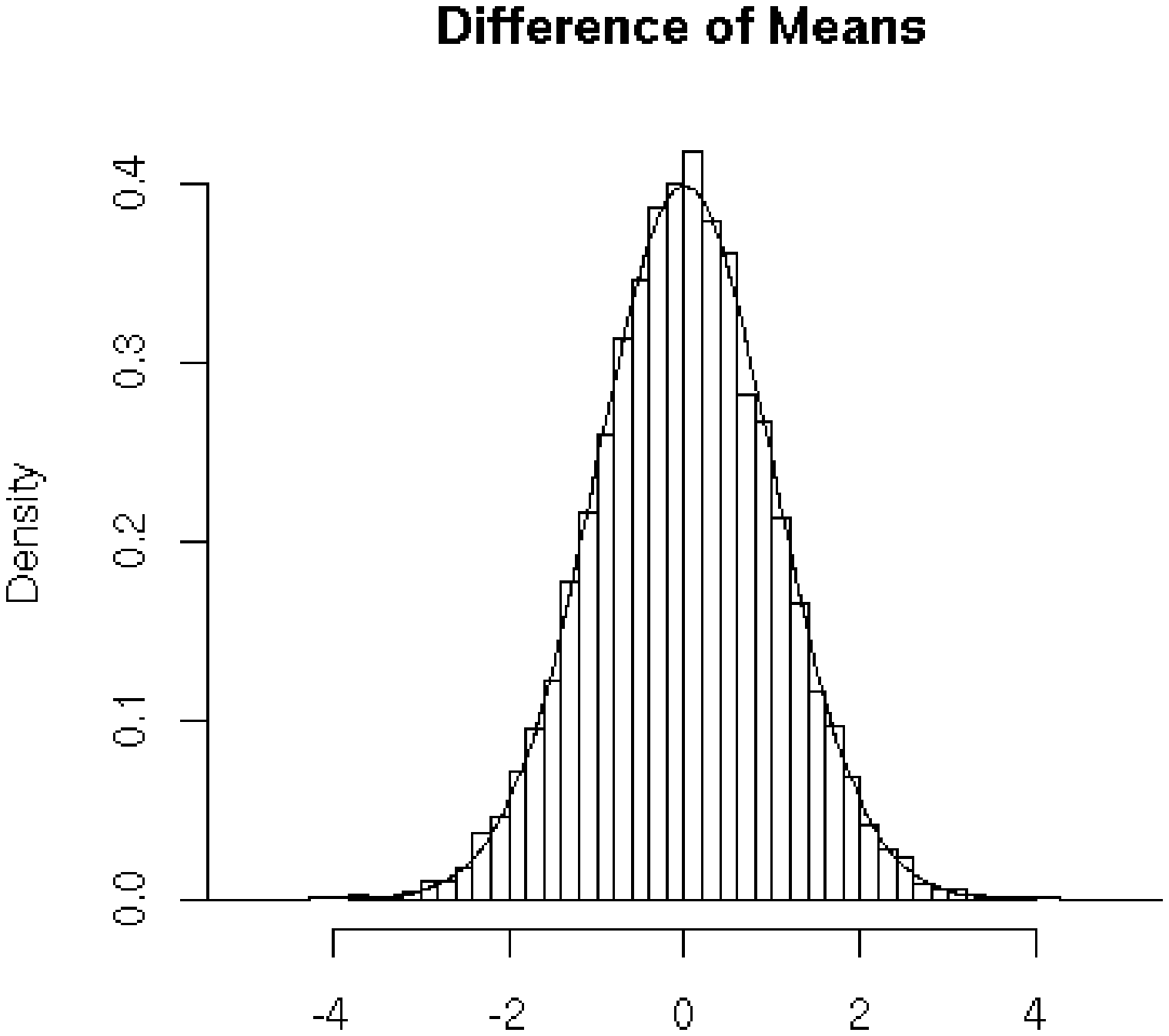}& \includegraphics[width=7cm,height=7cm]{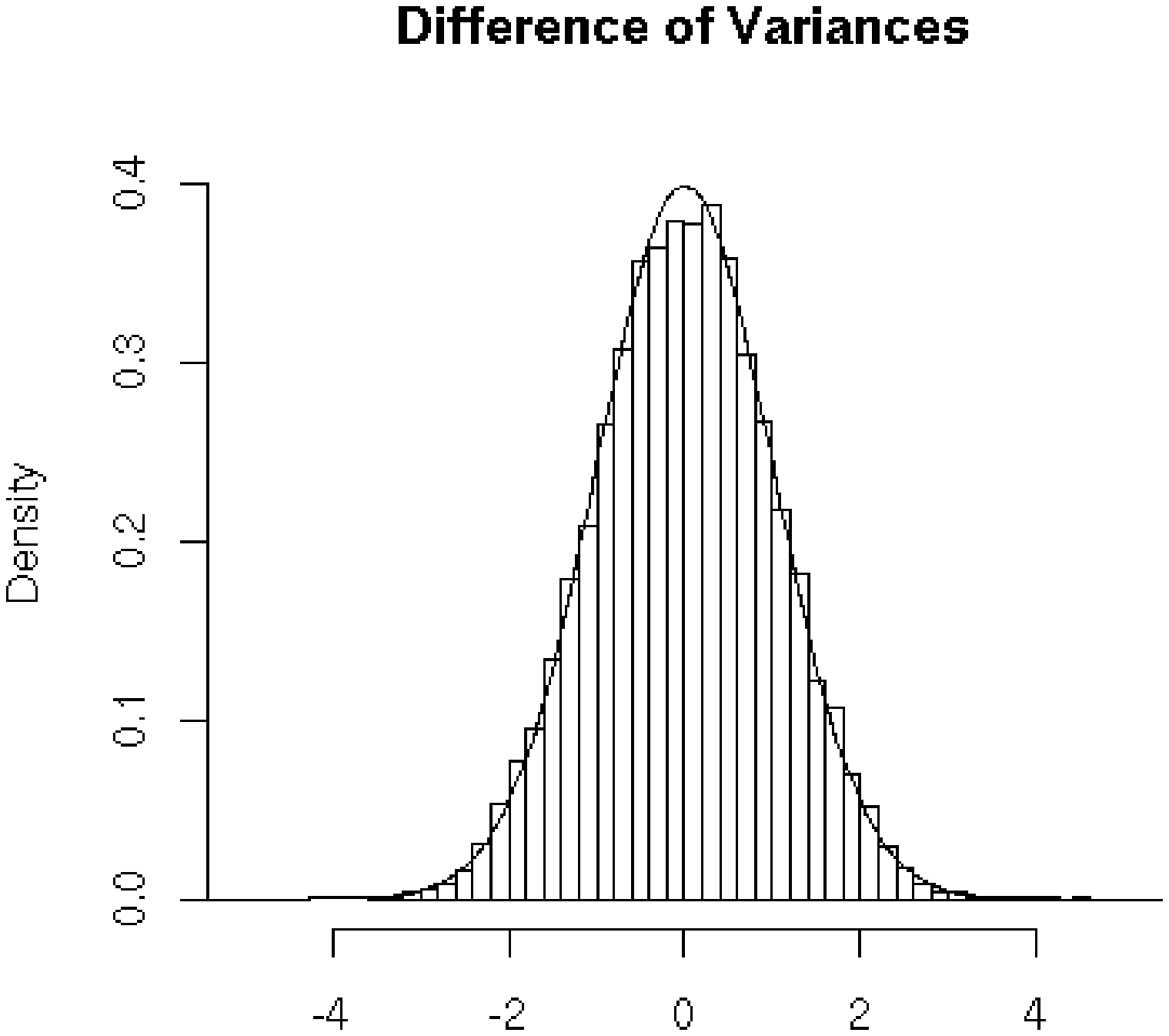} \\
\includegraphics[width=7cm,height=7cm]{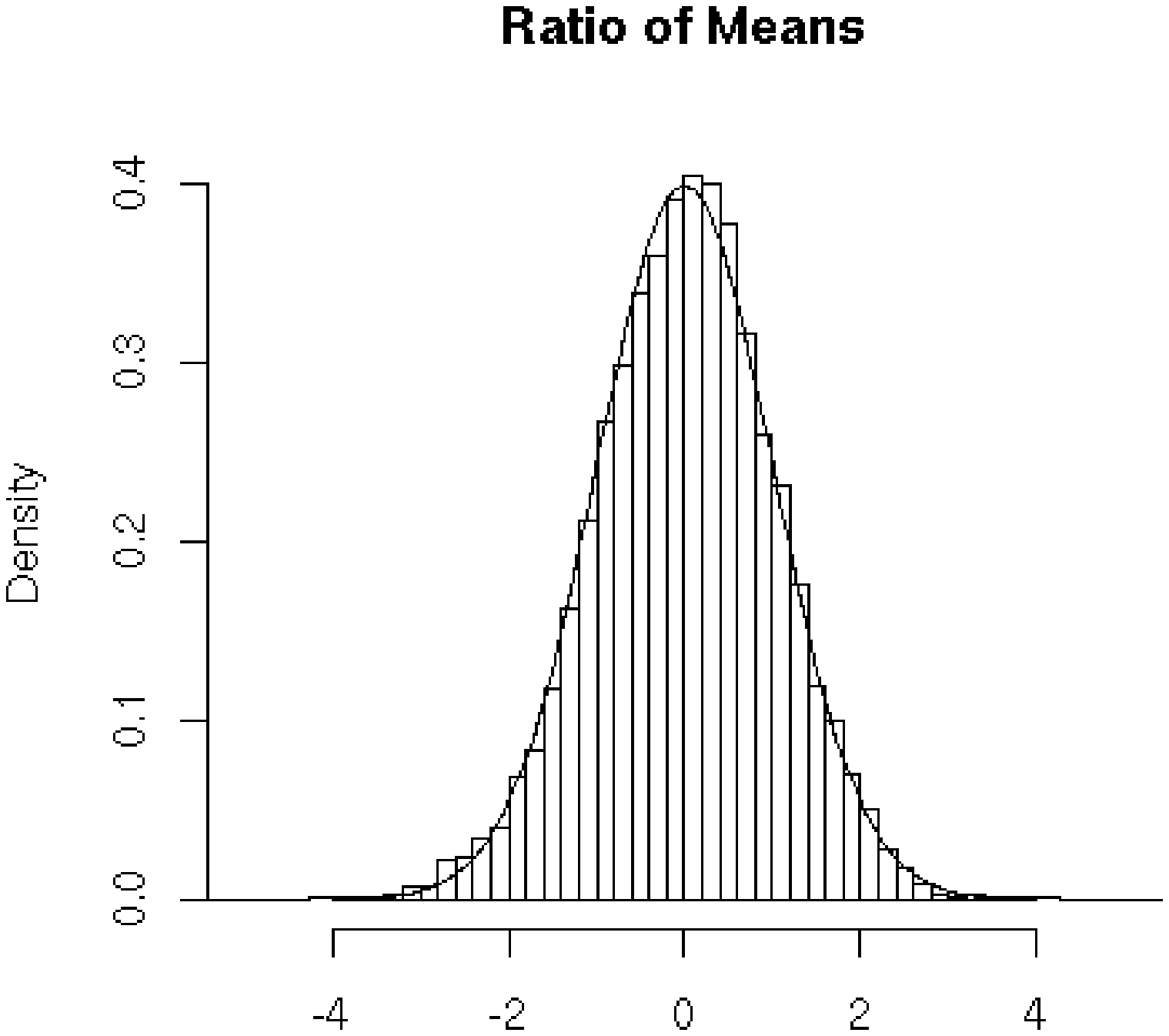}& \includegraphics[width=7cm,height=7cm]{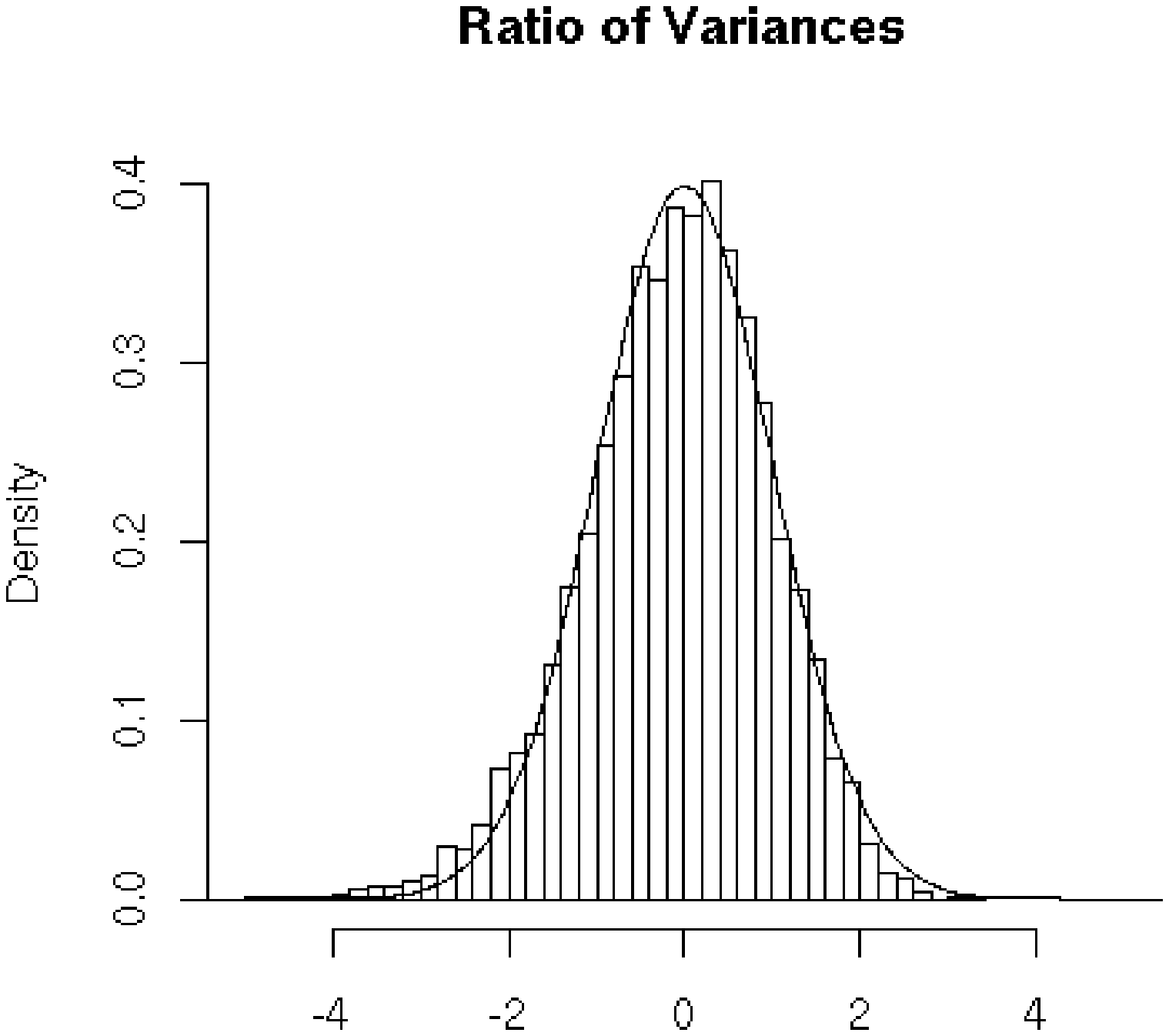} \\
\end{tabular}
\caption{Histograms of $m=10000$ replications of $\frac{\Est{\theta}{Y} - \theta}{\Est{\sigma_{\widehat{\theta}}}{Y}} $ for $\theta=\mu$, $\sigma^2$, $d_\mu$, $d_{\sigma^2}$, $r_\mu$ and $r_{\sigma^2}$. The simulation has been done as follows: $n=n^{(1)}=n^{(2)}=500$, $Y^{(1)}\leadsto \chi^2(5)$, $Y^{(2)}\leadsto \chi^2(5)$.} \label{fig-chi2}
\end{figure}

\begin{figure}
\begin{tabular}{cc}
\includegraphics[width=7cm,height=7cm]{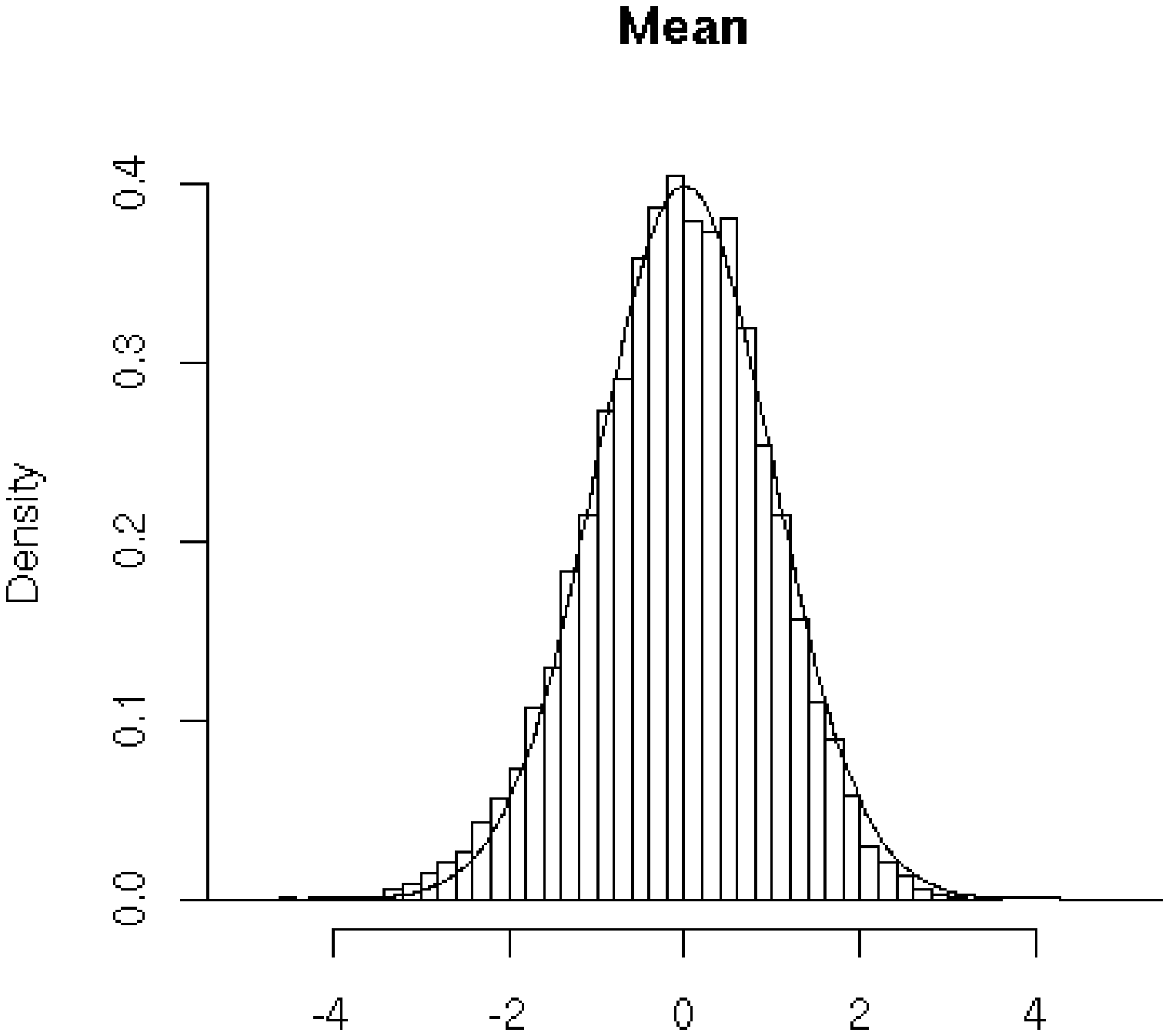}& \includegraphics[width=7cm,height=7cm]{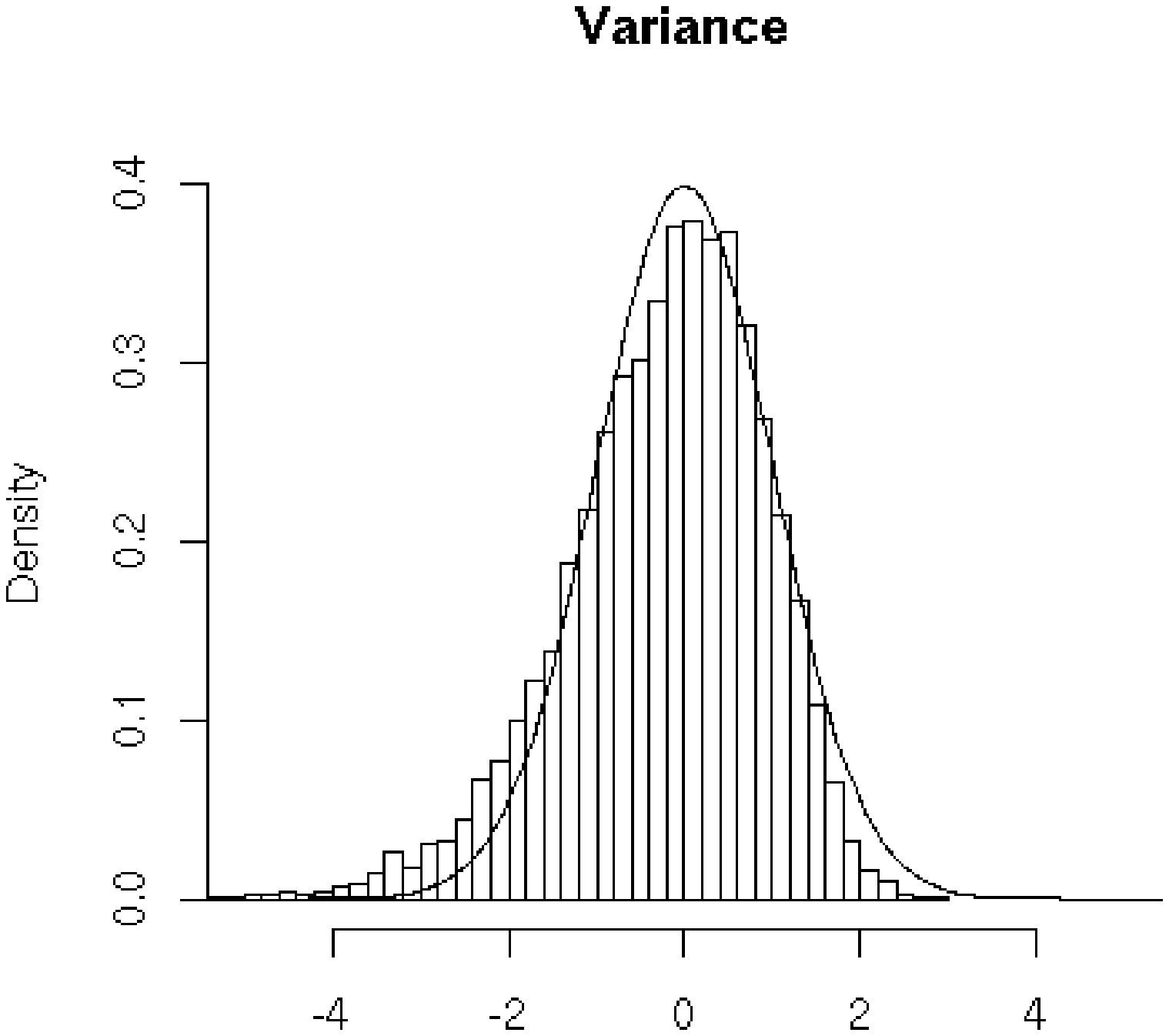} \\
\includegraphics[width=7cm,height=7cm]{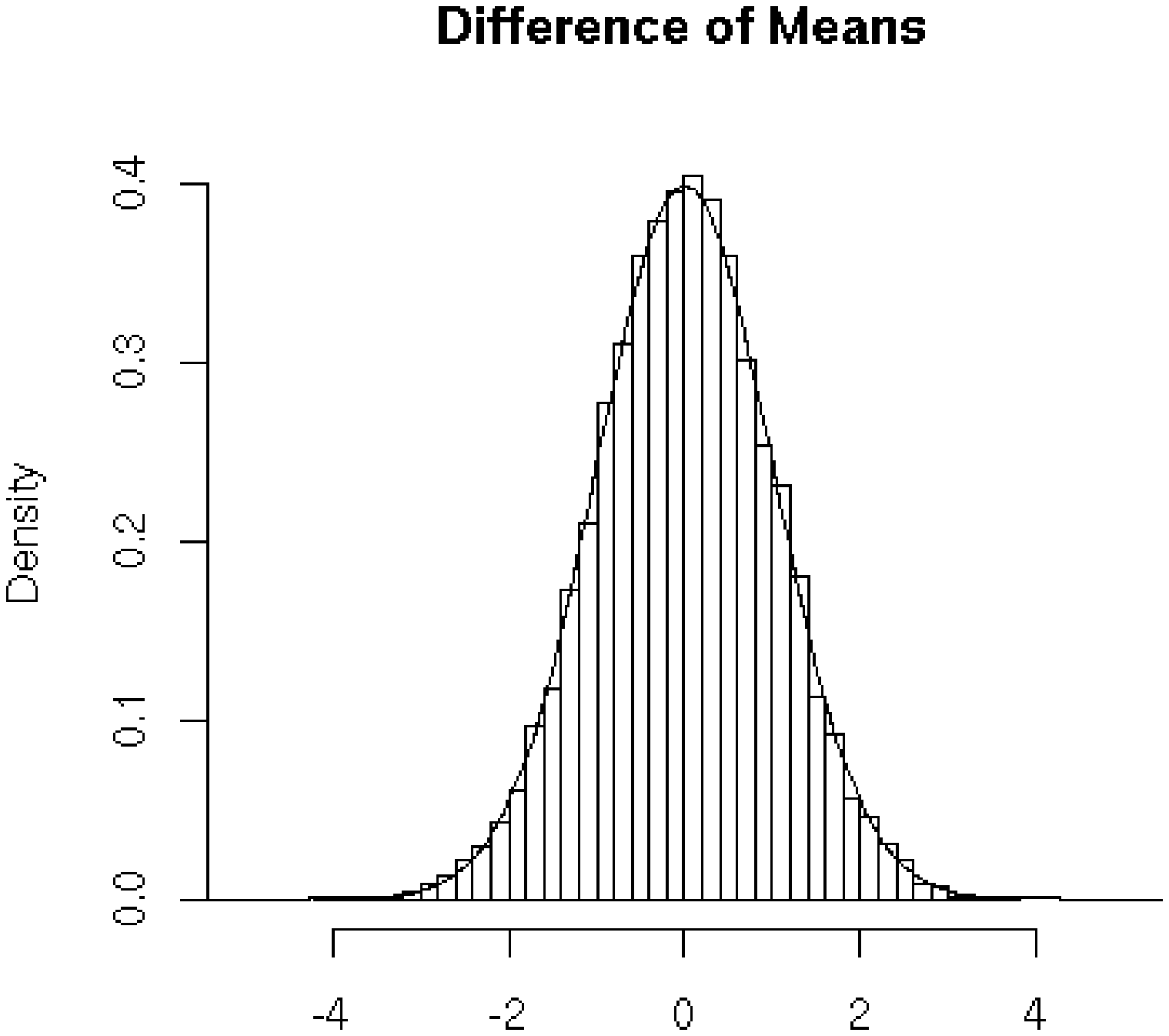}& \includegraphics[width=7cm,height=7cm]{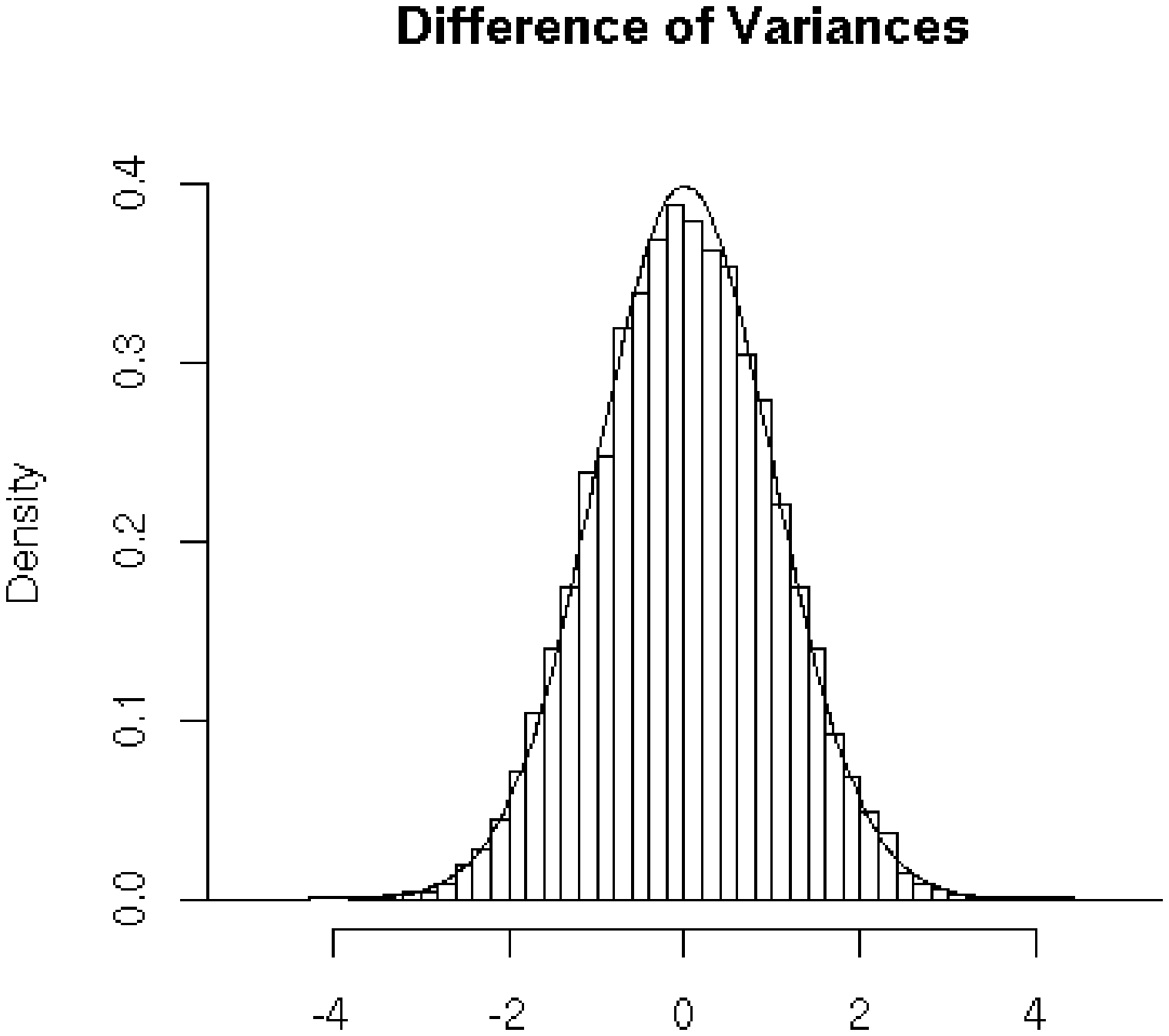} \\
\includegraphics[width=7cm,height=7cm]{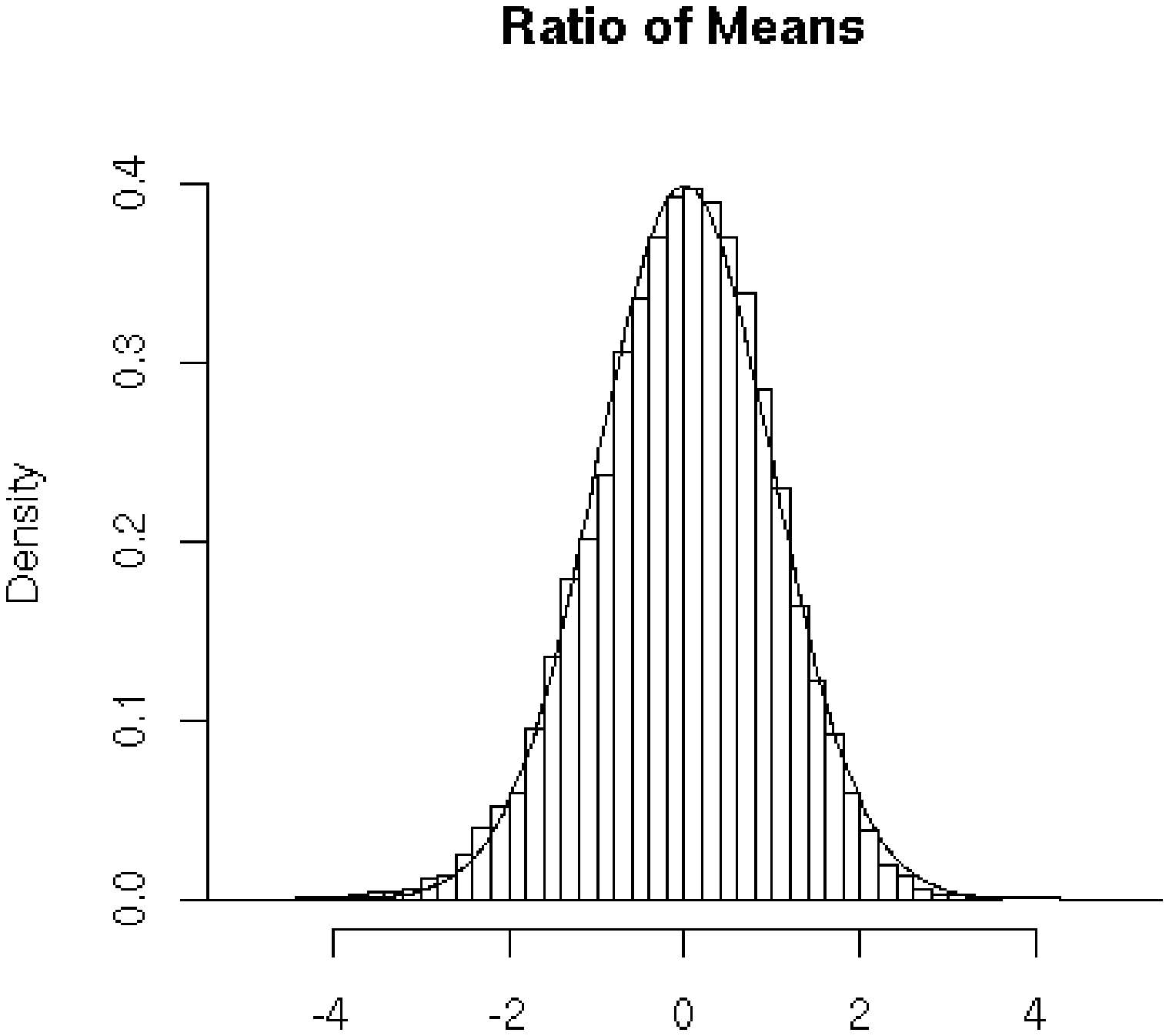}& \includegraphics[width=7cm,height=7cm]{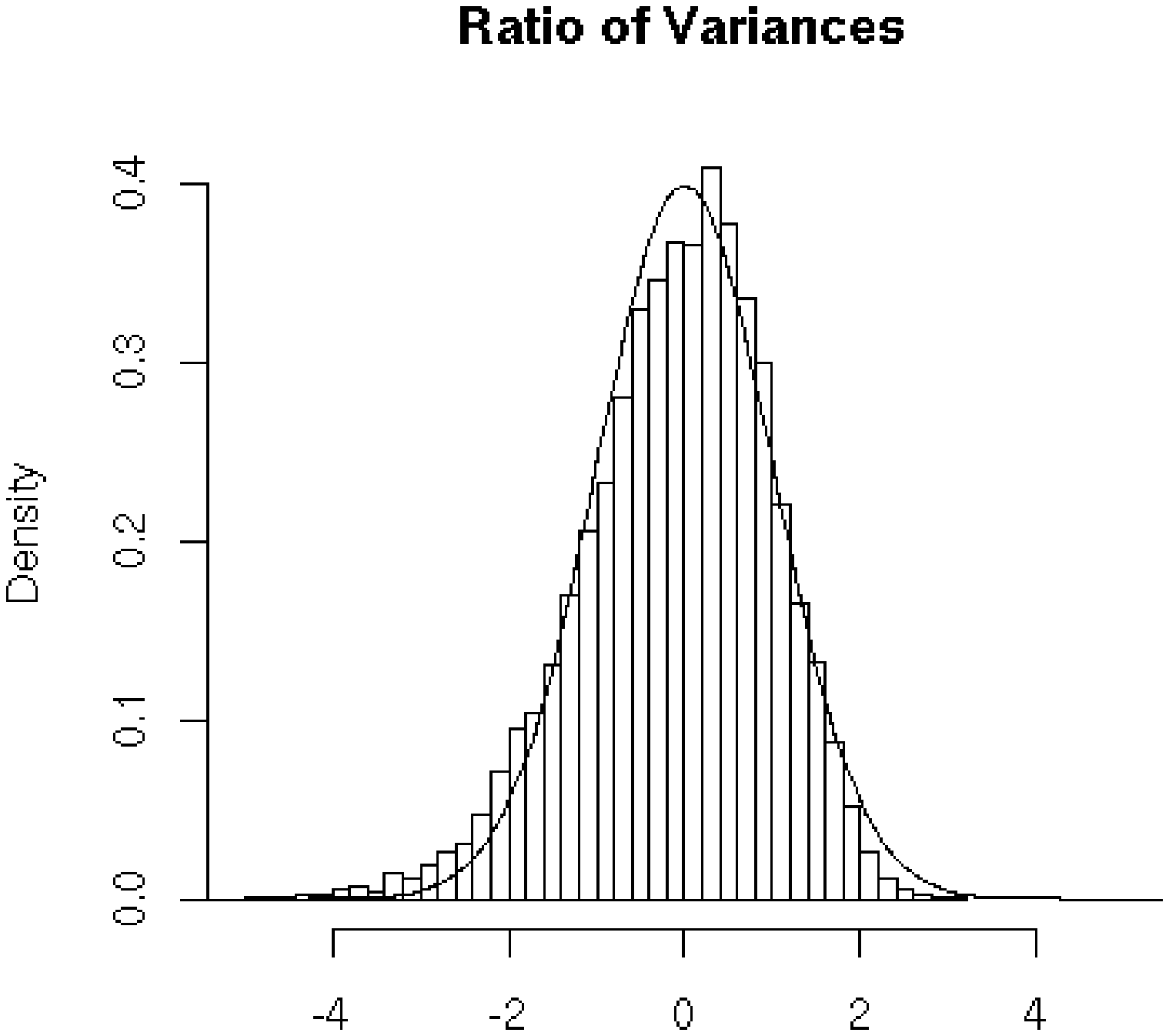} \\
\end{tabular}
\caption{Histograms of $m=10000$ replications of $\frac{\Est{\theta}{Y} - \theta}{\Est{\sigma_{\widehat{\theta}}}{Y}} $ for $\theta=\mu$, $\sigma^2$, $d_\mu$, $d_{\sigma^2}$, $r_\mu$ and $r_{\sigma^2}$. The simulation has been done as follows: $n=n^{(1)}=n^{(2)}=500$, $Y^{(1)}\leadsto \mathcal{E}(1)$, $Y^{(2)}\leadsto \mathcal{E}(1)$.} \label{fig-exp}
\end{figure}

\begin{figure}
\begin{tabular}{cc}
\includegraphics[width=7cm,height=7cm]{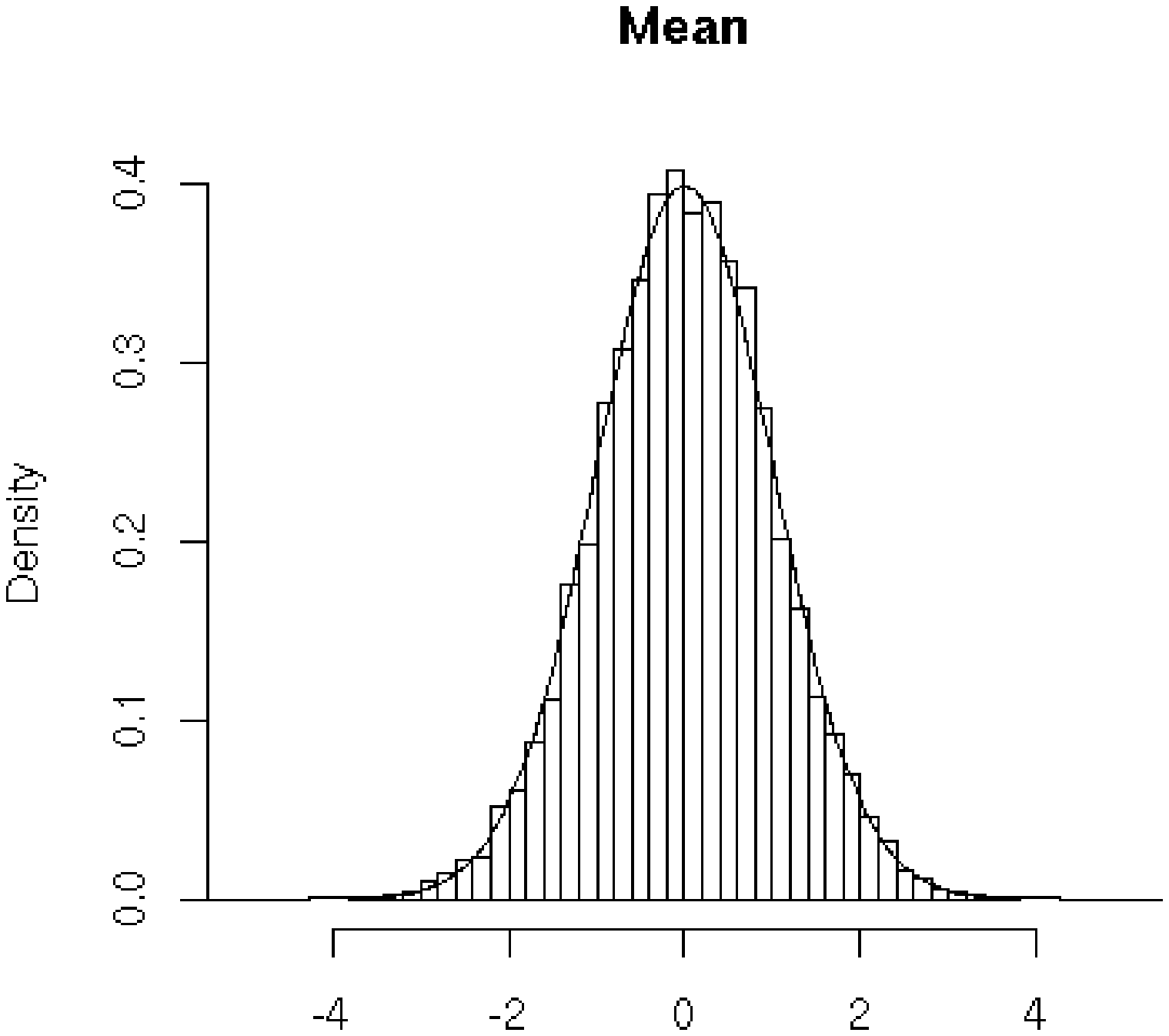}& \includegraphics[width=7cm,height=7cm]{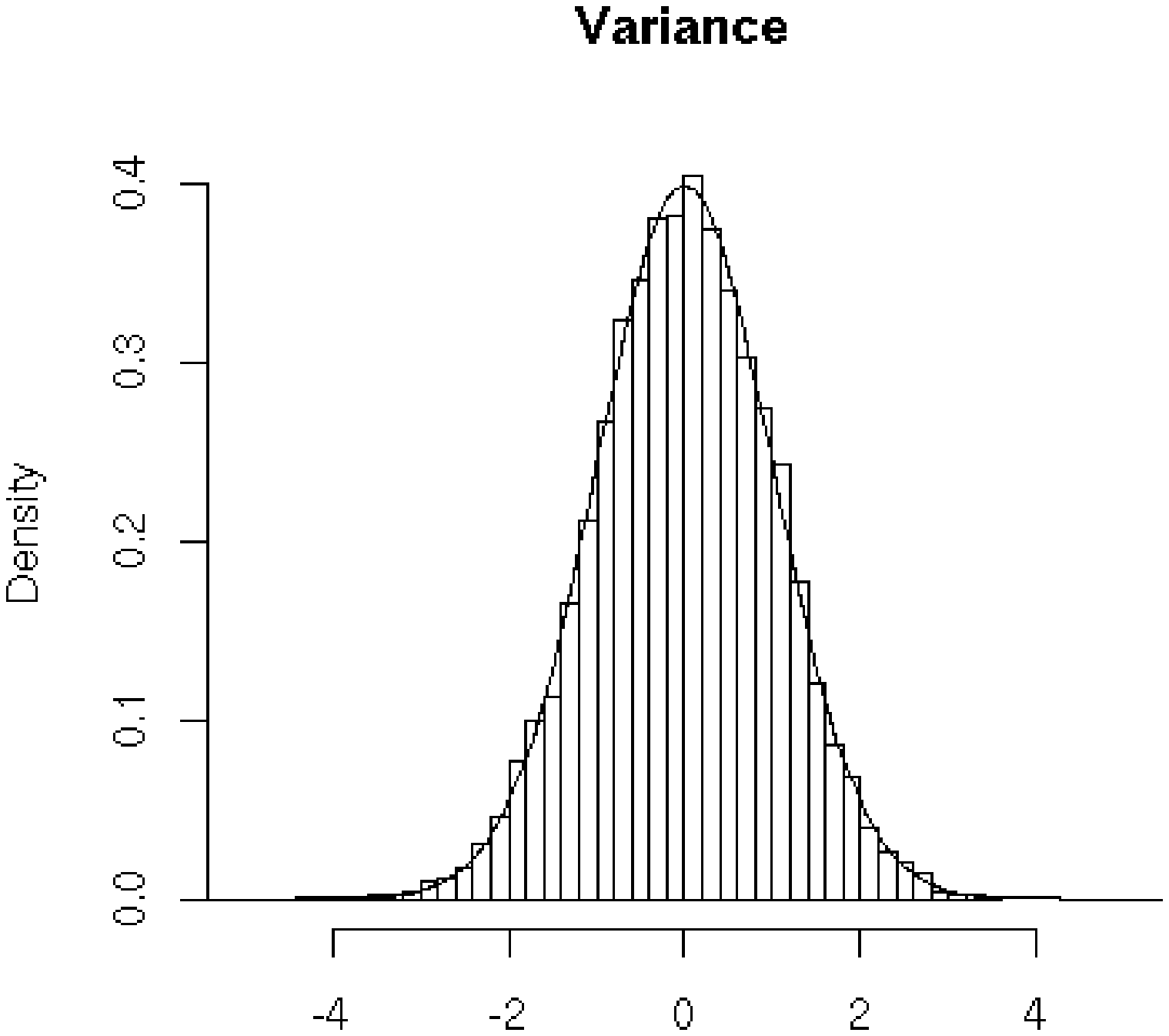} \\
\includegraphics[width=7cm,height=7cm]{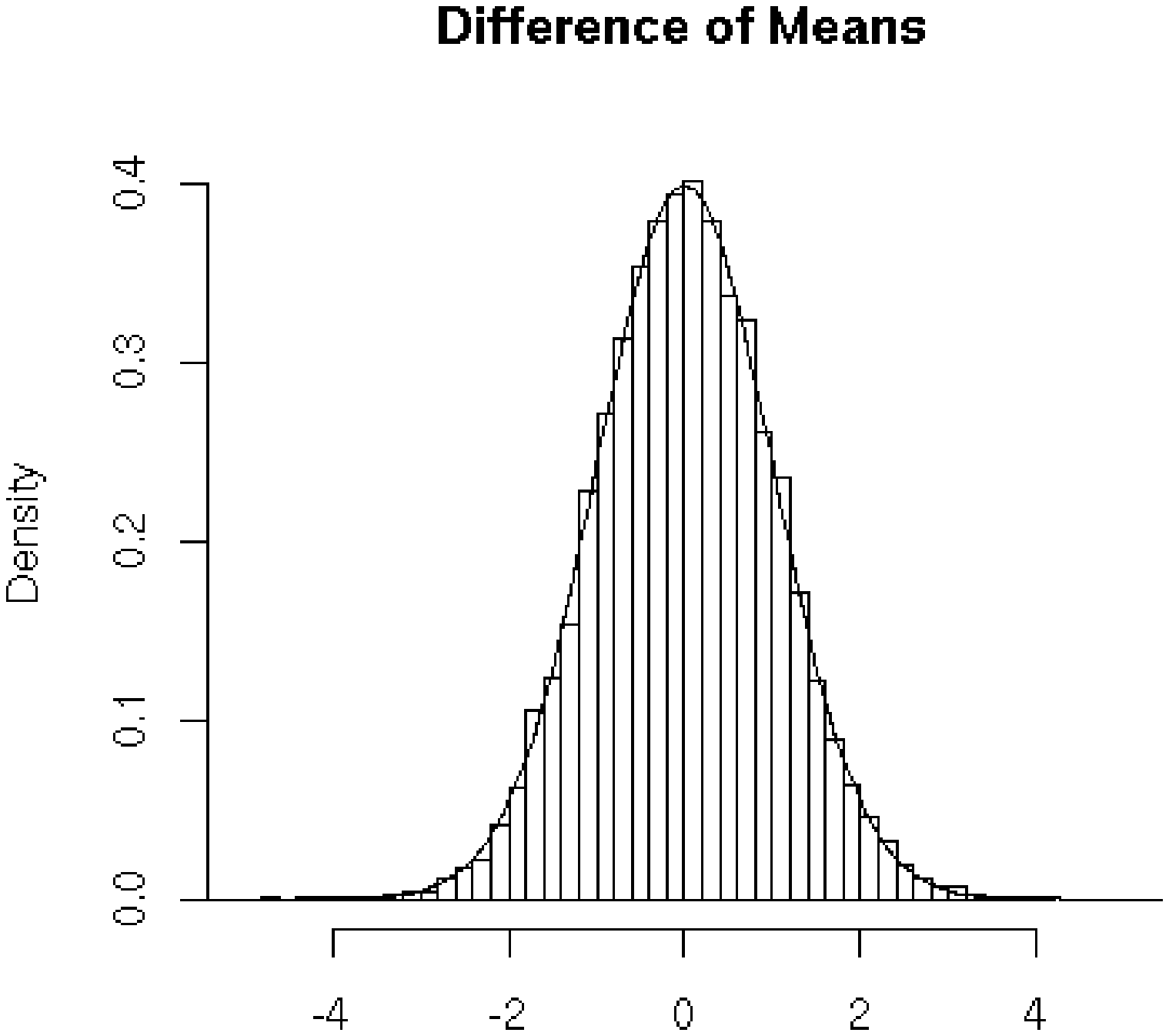}& \includegraphics[width=7cm,height=7cm]{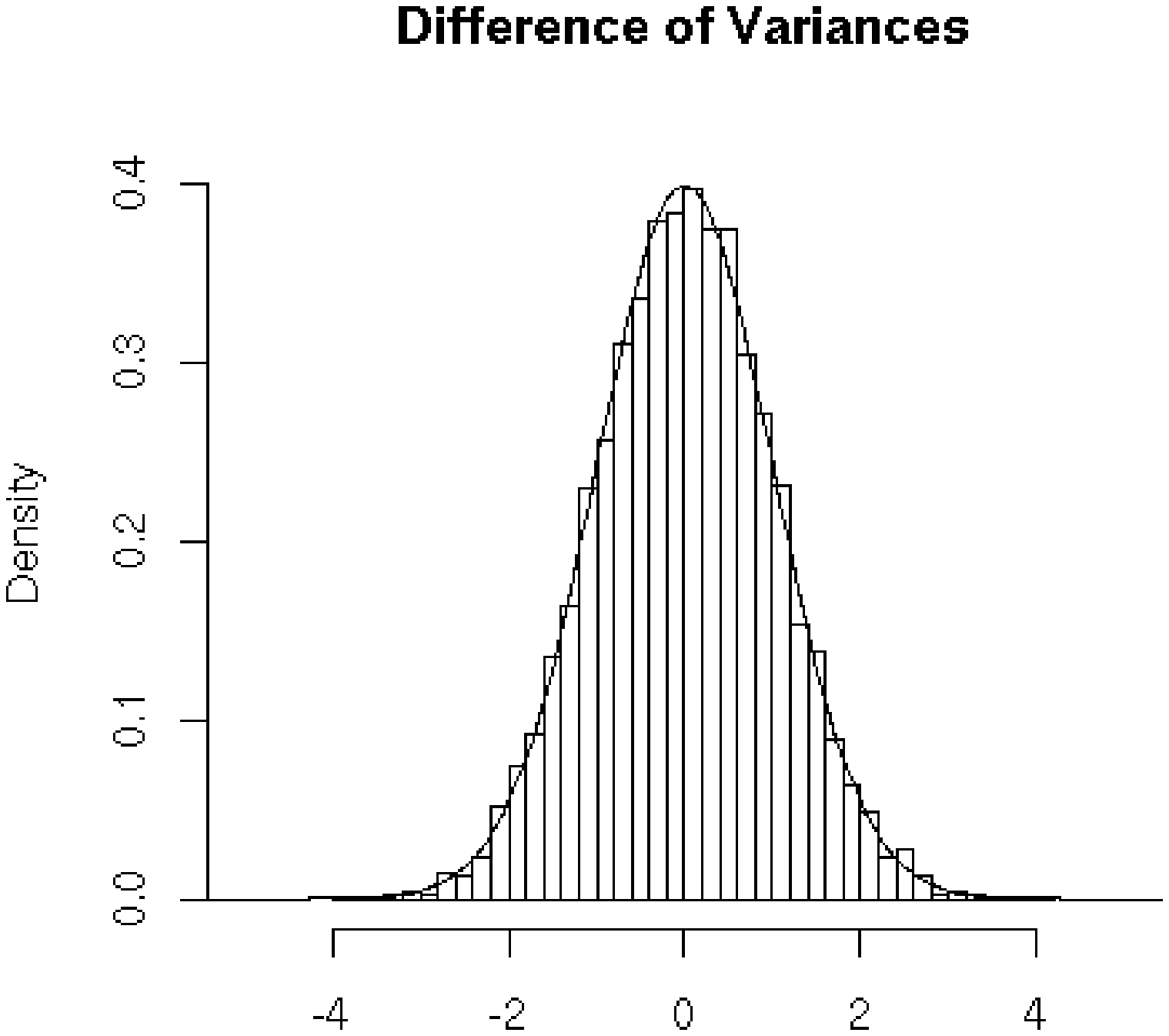} \\
\includegraphics[width=7cm,height=7cm]{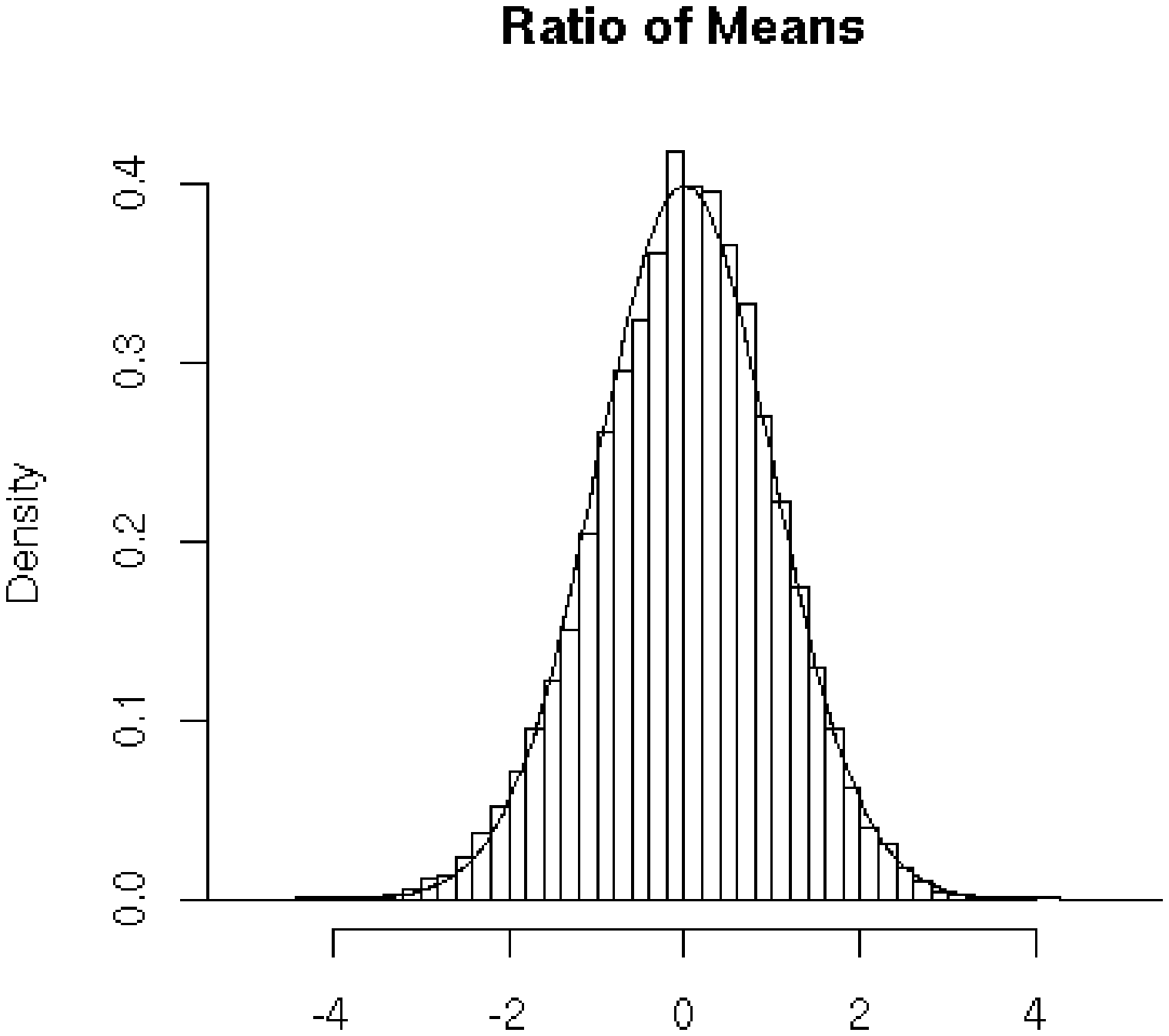}& \includegraphics[width=7cm,height=7cm]{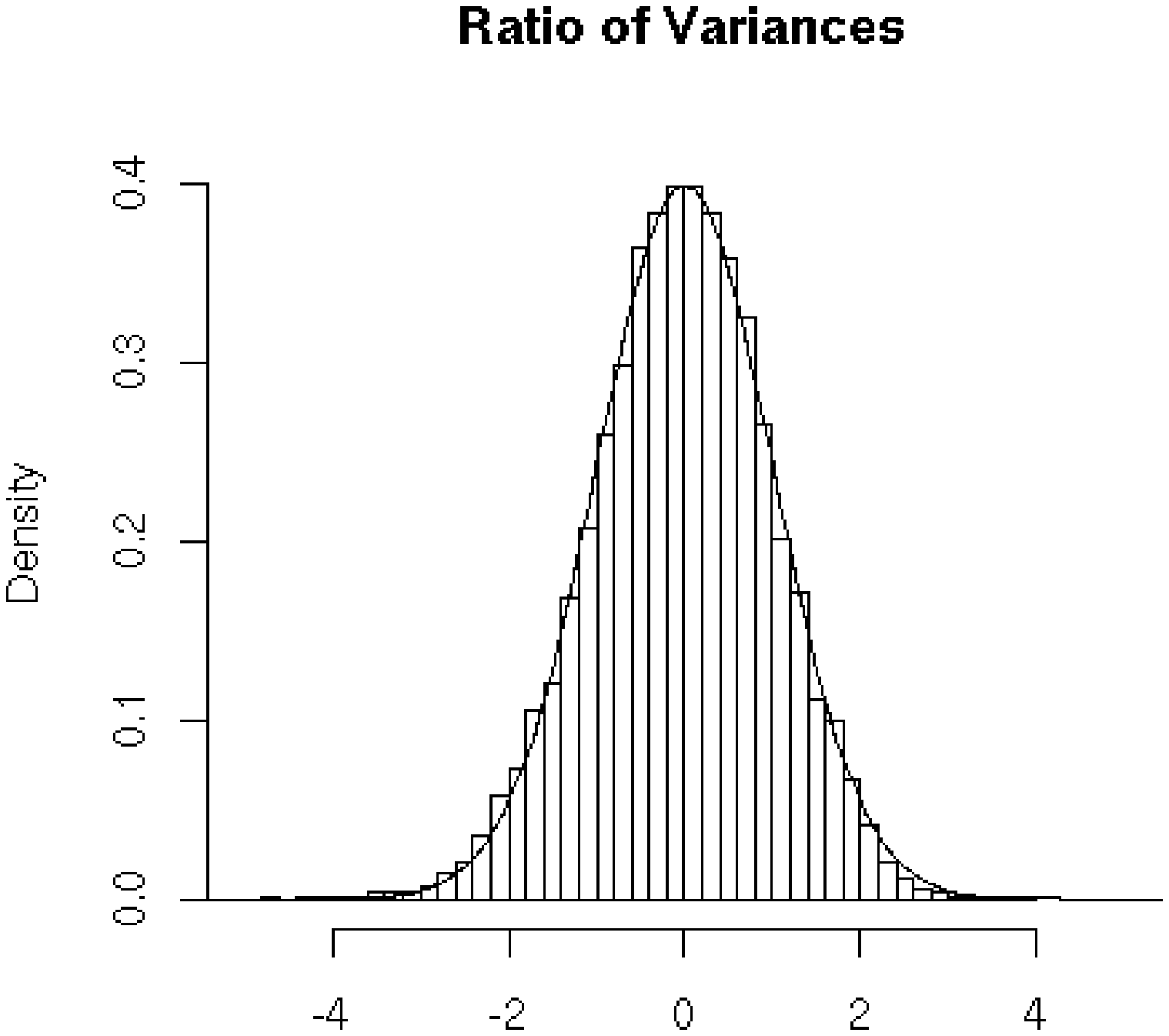} \\
\end{tabular}
\caption{Histograms of $m=10000$ replications of $\frac{\Est{\theta}{Y} - \theta}{\Est{\sigma_{\widehat{\theta}}}{Y}} $ for $\theta=\mu$, $\sigma^2$, $d_\mu$, $d_{\sigma^2}$, $r_\mu$ and $r_{\sigma^2}$. The simulation has been done as follows: $n=n^{(1)}=n^{(2)}=500$, $Y^{(1)}\leadsto \mathcal{U}([0,5])$, $Y^{(2)}\leadsto \mathcal{E}(0.5)$.} \label{fig-unif}
\end{figure}

\begin{remark}
The figures Fig.~\ref{fig-chi2}, Fig.~\ref{fig-exp} and Fig.~\ref{fig-unif} allow the reader to illustrate the mathematical result~(\ref{eq-tcl}).
\end{remark}

\begin{remark}
The asymptotic result~(\ref{eq-tcl}) allows us to easily construct statistical hypothesis tests and confidence intervals, see {\it e.g.} \cite{Casella90} p.~385 for development.
\end{remark}

\begin{remark}
The alternative hypothesis $\mathbf{H_1}$ comparing the ratio of means or variances to some reference value $r_0$ may be expressed in terms of the comparison of the weighted differences $d_{\mu}:=\mu^{(1)}-r_0\mu^{(2)}$ or $d_{\sigma^2}:=\sigma^2_{(1)}-r_0\sigma^2_{(2)}$ to  0, the value of $\rho$ being fixed to $r_0$.

\end{remark}

\section{Using {asympTest}} \label{sec-pkg}

The \proglang{R} package \pkg{asympTest} consists of a main function \texttt{asymp.test} and six auxiliary ones designed to compute standard errors of estimates of different parameters, see Tab.~\ref{tab-se}. The auxiliary functions should not be very useful for the user, except if he/she wants to compute himself/herself the confidence interval. The function \texttt{asymp.test} has been written in the same spirit as the \proglang{R} functions \texttt{t.test} or \texttt{var.test}. The arguments of \texttt{asymp.test}, its value and the  resulting outputs are inspired from the ones of \texttt{t.test} or \texttt{var.test}. In particular, the function \texttt{asympt.test} returns an object of class ``htest'' (which is the general class of test objects in \proglang{R}, see \cite{RTeam04}).

The main arguments of the function \texttt{asymp.test} are:
\begin{itemize}
\item \texttt{x}: vector of data values.
\item \texttt{y}: optional vector of data values.
\item \texttt{parameter}: parameter under testing, must be one of ``mean'', ``var'', ``dMean'', ``dVar'', ``rMean'', ``rVar''.
\item \texttt{alternative}: alternative hypothesis, must be one of  ``two.sided" (default), ``greater'' or ``less''.
\item \texttt{reference}: reference value of the parameter under the null hypothesis.
\item \texttt{conf.level}: confidence level of the interval (default is 0.95). The type-one error is then fixed to 1-\texttt{conf.level}.
\item \texttt{rho}: optional parameter (only used for parameters "dMean" and "dVar") for penalization (or enhancement) of the contribution of the second parameter.
\end{itemize}
The user may only specify the first letters of the parameter or alternative.

In order to illustrate this function, let us consider the \texttt{iris} data available in \proglang{R}. This famous (Fisher's or Anderson's \cite{Fisher36,Anderson35}) data set gives the measurements (in centimeters) of the four variables sepal length and width, and petal length and width, for 50 flowers from each species of iris: setosa, versicolor, and virginica.

\begin{Verbatim}[frame=leftline,fontfamily=tt,fontshape=n,numbers=left]
> data(iris)
> attach(iris)
> names(iris)
[1] "Sepal.Length" "Sepal.Width"  "Petal.Length" "Petal.Width"  "Species"     
> levels(iris$Species)
[1] "setosa"     "versicolor" "virginica"
\end{Verbatim}

The following table presents the p-values of all the Shapiro-Wilk normality tests for the different variables and the three species.

\begin{table}[ht]
\begin{center}
\begin{tabular}{rrrrr}
  \hline
 & Sepal.Length & Sepal.Width & Petal.Length & Petal.Width \\
  \hline
setosa & 0.4595 & 0.2715 & 0.0548 & 0.0000 \\
  versicolor & 0.4647 & 0.3380 & 0.1585 & 0.0273 \\
  virginica & 0.2583 & 0.1809 & 0.1098 & 0.0870 \\
   \hline
\end{tabular}
\end{center}
\end{table}

Let us concentrate on the variable Petal.Width for which the Gaussian assumption seems to be wrong for each one of the three species. 
The empirical means and variances are given below:
\begin{Verbatim}[frame=leftline,fontfamily=tt,fontshape=n,numbers=left]
> by(Petal.Width,Species,function(e) c(mean=mean(e),var=var(e)))
Species: setosa
      mean        var 
0.24600000 0.01110612 
------------------------------------------------------------ 
Species: versicolor
      mean        var 
1.32600000 0.03910612 
------------------------------------------------------------ 
Species: virginica
      mean        var 
2.02600000 0.07543265
\end{Verbatim}

Is the mean petal width of setosa species less than $0.5$~?

\begin{Verbatim}[frame=leftline,fontfamily=tt,fontshape=n,numbers=left]
> require(asympTest)
> asymp.test(Petal.Width[Species=="setosa"],par="mean",alt="l",ref=0.5)

	One-sample asymptotic mean test

data:  Petal.Width[Species == "setosa"] 
statistic = -17.0427, p-value < 2.2e-16
alternative hypothesis: true mean is less than 0.5 
95 percent confidence interval:
      -Inf 0.2705145 
sample estimates:
 mean 
0.246
\end{Verbatim}

Is the mean petal width of virginica species larger than the versicolor one~?

\begin{Verbatim}[frame=leftline,fontfamily=tt,fontshape=n,numbers=left]
> asymp.test(Petal.Width[Species=="virginica"],
+ Petal.Width[Species=="versicolor"],"dMean","g",0)

	Two-sample asymptotic difference of means test

data:  Petal.Width[Species == "virginica"] and Petal.Width[Species == "versicolor"] 
statistic = 14.6254, p-value < 2.2e-16
alternative hypothesis: true difference of means is greater than 0 
95 percent confidence interval:
 0.621274      Inf 
sample estimates:
difference of means 
                0.7
\end{Verbatim}

Is the mean petal width of virginica species 4 times larger than the setosa one~?

\begin{Verbatim}[frame=leftline,fontfamily=tt,fontshape=n,numbers=left]
> asymp.test(Petal.Width[Species=="virginica"],
+ Petal.Width[Species=="setosa"],"rMean","g",4)

	Two-sample asymptotic ratio of means test

data:  Petal.Width[Species == "virginica"] and Petal.Width[Species == "setosa"] 
statistic = 8.0936, p-value = 3.331e-16
alternative hypothesis: true ratio of means is greater than 4 
95 percent confidence interval:
 7.374946      Inf 
sample estimates:
ratio of means 
      8.235772
\end{Verbatim}

This may also be done via a  difference of weighted means test.

\begin{Verbatim}[frame=leftline,fontfamily=tt,fontshape=n,numbers=left]
> asymp.test(Petal.Width[Species=="virginica"],
+ Petal.Width[Species=="setosa"],"dMean","g",0,rho=4)

	Two-sample asymptotic difference of (weighted) means test

data:  Petal.Width[Species == "virginica"] and Petal.Width[Species == "setosa"] 
statistic = 14.6447, p-value < 2.2e-16
alternative hypothesis: true difference of (weighted) means is greater than 0 
95 percent confidence interval:
 0.9249653       Inf 
sample estimates:
difference of (weighted) means 
                         1.042
\end{Verbatim}

\section{Type I error risks} \label{sec-comp}

\subsection{Comparison between classical and asymptotic variance tests}

In the context of large samples, two simulation studies are proposed in order to show the lack of reliability of the classical tests for variance parameters compared with the asymptotic tests studied in this paper. For each of the following examples, 10000 simulations of samples of size $n=1000$ have been performed.

\begin{enumerate}
\item Let us consider testing  $\mathbf{H_0}: \sigma^2=1$ versus $\mathbf{H_1}:\sigma^2<1$ with data sampled from distribution $\mathcal{E}(1)$  (i.e., under $\mathbf{H_0}$).

\begin{table}[htbp]
\begin{center}
\begin{tabular}{rrr}
  \hline
 & false & true \\
  \hline
false & 0.7901 & 0.0000 \\
  true & 0.1208 & 0.0891 \\
   \hline
\end{tabular}
\caption{Acceptance of $\mathbf{H_1}$ for the chi-square test (rows) versus the asymptotic test (columns)}
\label{tab-exp}
\end{center}
\end{table}

In the case where $\alpha=5\%$, the probability of accepting the alternative hypothesis is 8.91\% for the asymptotic test and  20.99\% for the chi-square test.
\item Let us consider test $\mathbf{H_0}: \sigma^2_{(1)}=\sigma_{(2)}^2$ versus $\mathbf{H_1}:\sigma^2_{(1)}\neq\sigma_{(2)}^2$ with data sampled from distribution $\mathcal{U}([0,5])$ for both samples (i.e., under $\mathbf{H_0}$).

\begin{table}[htbp]
\begin{center}
\begin{tabular}{rrr}
  \hline
 & false & true \\
  \hline
false & 0.9498 & 0.0481 \\
  true & 0.0000 & 0.0021 \\
   \hline
\end{tabular}
\caption{Acceptance of $\mathbf{H_1}$ for the Fisher test (rows) versus the asymptotic test (columns)}
\label{tab-unif}
\end{center}
\end{table}

In the case where $\alpha=5\%$, the probability of accepting the alternative hypothesis is 5.02\% for the asymptotic test and  0.21\% for the Fisher test.

\end{enumerate}
In both cases, the probabilities for Type I errors are worse for the classical tests than for the corresponding well-suited asymptotic tests. This is a direct consequence of the previous results summarized by figure Fig.~\ref{fig-gauss}.

\subsection{Back to the example of the introduction}
\noindent Now, one may wonder what the consequences of these previous results are  for pratical purposes. Let us consider again the example presented in the introduction. In order to illustrate the two-samples case, we propose a second example.
In the following \texttt{R} outputs, the data of the first  (resp. second) example are denoted by \texttt{y} (resp. \texttt{y1} and \texttt{y2}). These samples have a size n=1000 and their empirical distributions, proposed below, do not seem to fit normal distributions.

\begin{tabular}{lll}
\includegraphics[width=5cm]{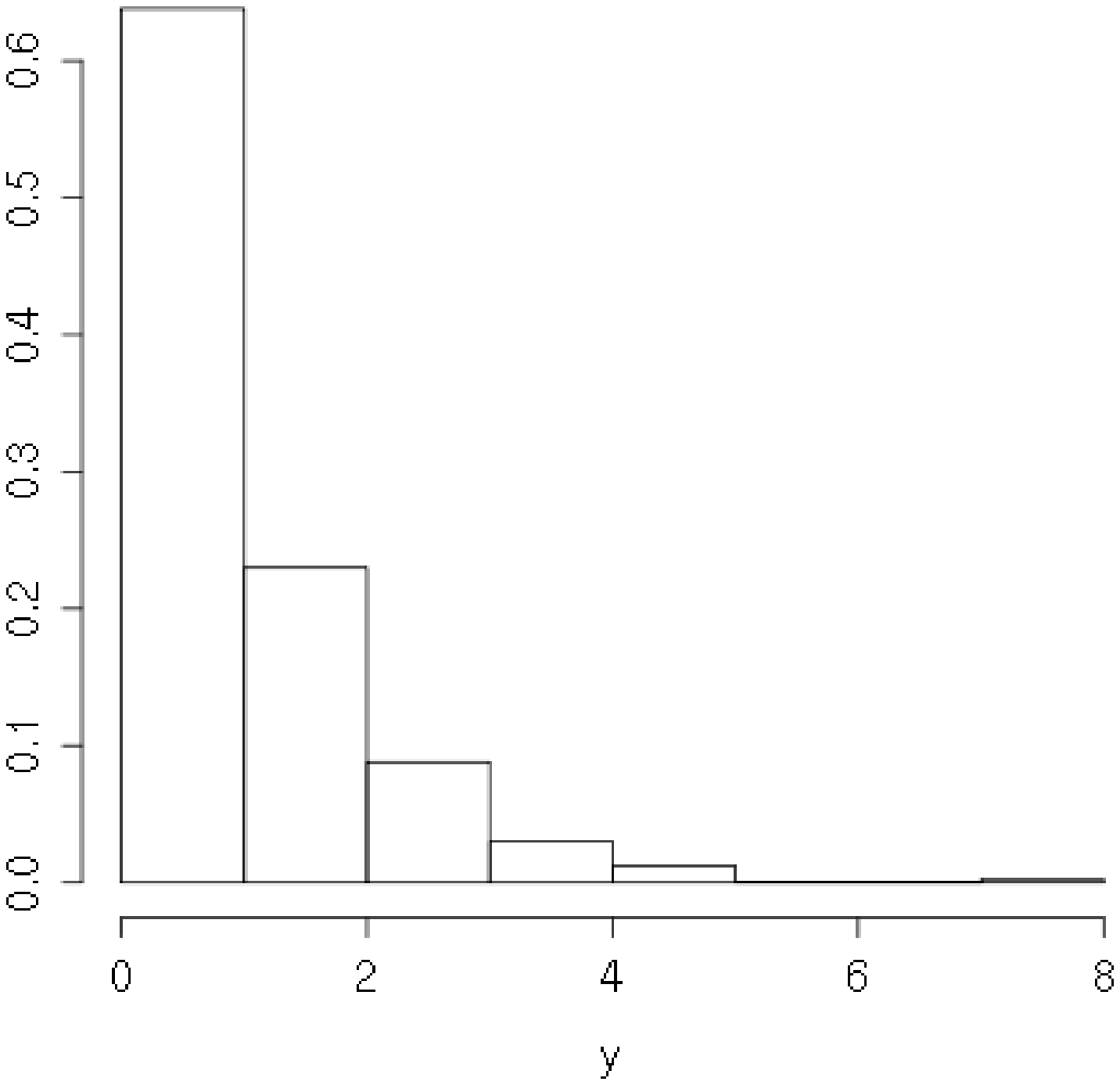} &\includegraphics[width=5cm]{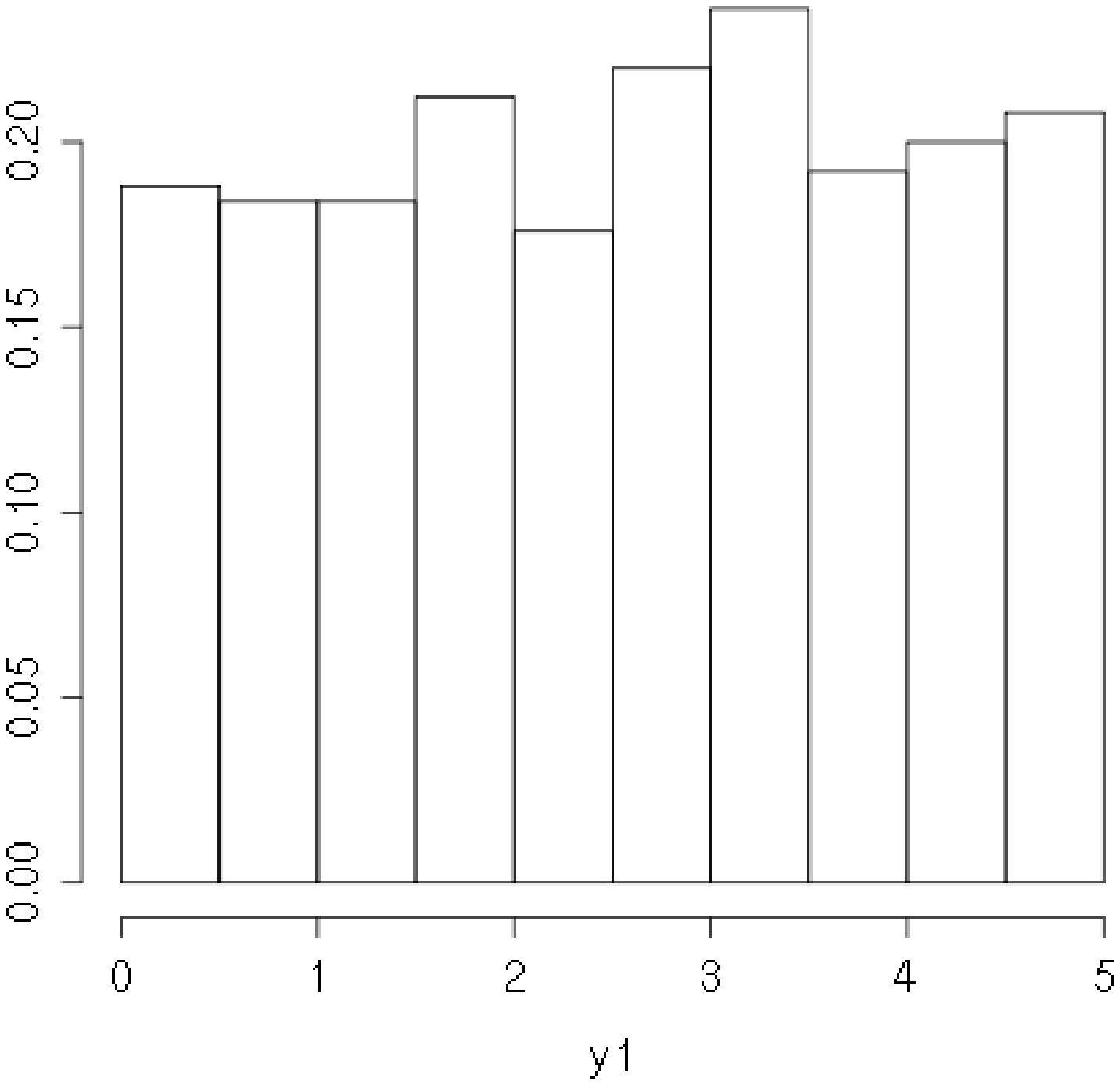}&\includegraphics[width=5cm]{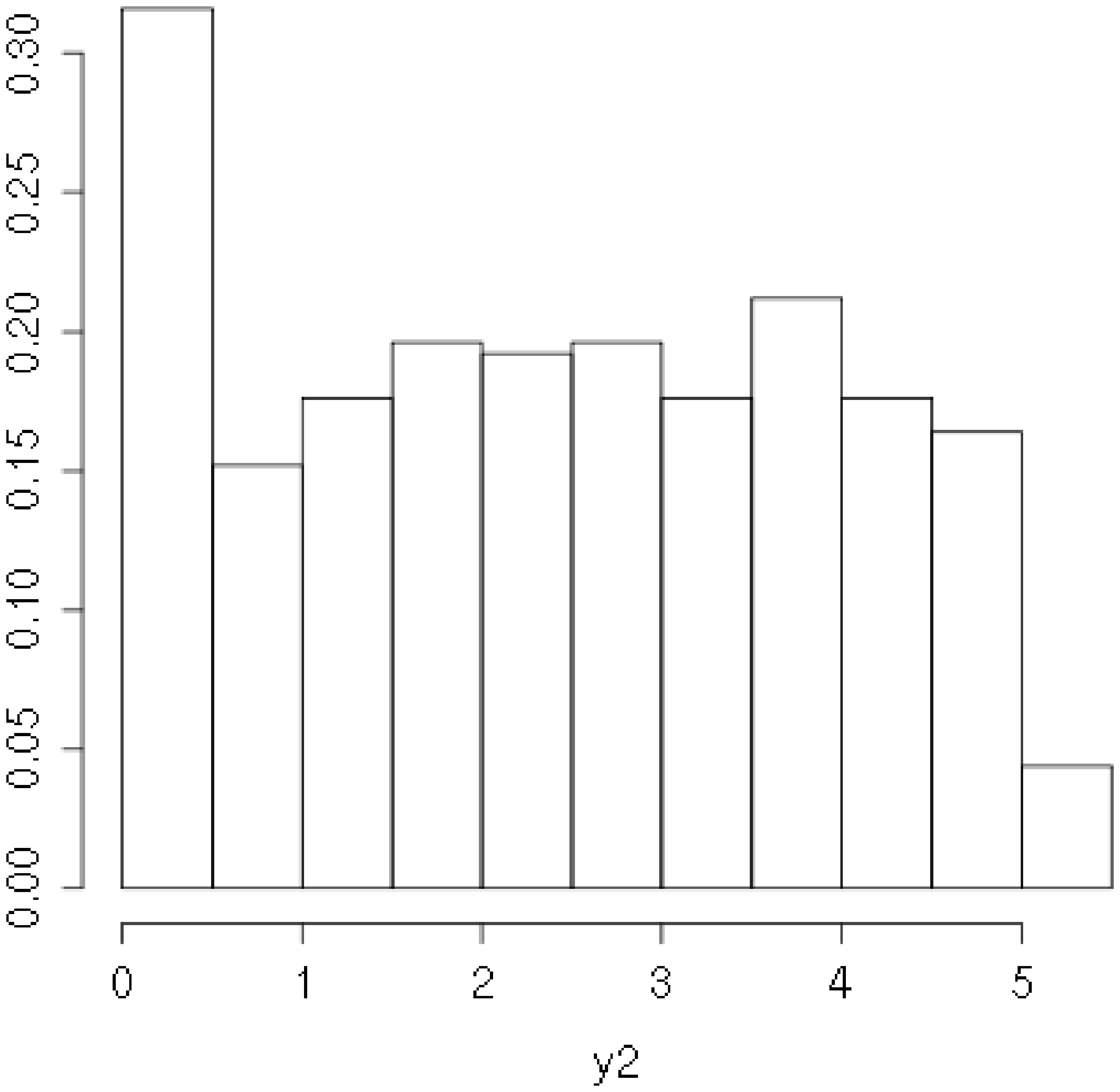}\\
\end{tabular}

The following output provides the p-value of the chi-square test for the first example.
\begin{Verbatim}[frame=leftline,fontfamily=tt,fontshape=n,numbers=left]
> pchisq((length(y)-1)*var(y)/1,length(y)-1)
[1] 0.04785152
\end{Verbatim}

Due to the apparent non normality of the data, one may prefer to apply the corresponding asymptotic test:
\begin{Verbatim}[frame=leftline,fontfamily=tt,fontshape=n,numbers=left]
> asymp.test(y,par="var",alt="l",ref=1)

	One-sample asymptotic variance test

data:  y 
statistic = -0.9771, p-value = 0.1643
alternative hypothesis: true variance is less than 1 
95 percent confidence interval:
     -Inf 1.070430 
sample estimates:
 variance 
0.8969455
\end{Verbatim}

The two decisions do not match, but since the empirical variance is 0.8969455, one may think that $\sigma^2$ is slightly inferior to 1. In which  case, we have to be cautious because our sample might be of the same kind as the 12.08\% of Table~\ref{tab-exp}.

The following output provides the p-value of the Fisher test for the second example.
\begin{Verbatim}[frame=leftline,fontfamily=tt,fontshape=n,numbers=left]
> var.test(y1,y2)

	F test to compare two variances

data:  y1 and y2 
F = 0.8874, num df = 499, denom df = 499, p-value = 0.1825
alternative hypothesis: true ratio of variances is not equal to 1 
95 percent confidence interval:
 0.7444324 1.0578390 
sample estimates:
ratio of variances 
         0.8874061
\end{Verbatim}

Due to the apparent non normality of the data, one may prefer to apply the corresponding asymptotic test:
\begin{Verbatim}[frame=leftline,fontfamily=tt,fontshape=n,numbers=left]
> asymp.test(y1,y2,"dVar")

	Two-sample asymptotic difference of variances test

data:  y1 and y2 
statistic = -2.0925, p-value = 0.03639
alternative hypothesis: true difference of variances is not equal to 0 
95 percent confidence interval:
 -0.49988046 -0.01635305 
sample estimates:
difference of variances 
             -0.2581168
\end{Verbatim}

The two decisions do not match, but since the empirical variances are 2.034341 and 2.292458, one may think that $\sigma_{(1)}^2$ and $\sigma_{(2)}^2$ are  slightly different. In which case, we have to be cautious because our sample might be of the same kind as the 4.81\% of Table~\ref{tab-unif}.

\section{Discussion}\label{sec-discuss}

We have presented an \proglang{R} package implementing large sample tests for various parameters. The interesting point is that each test   statistic can be written in the same form, as follows: 
$$
\EcartEst{\theta} :=\frac{\Est{\theta}{Y} - \theta}{\Est{\sigma_{\widehat{\theta}}}{Y}}.
$$
This form clearly expresses the departure of the estimate from the true parameter, normalized by some quantity measuring the precision of the estimate. This approach is then attractive and easy to present. In the Gaussian framework, the one and two-sample t-tests follow this idea whereas the chi-square variance test and the Fisher test do not.  
One may wonder if it is possible to embed the classical Gaussian framework within this formalism.
 More precisely, for an estimate $\Est{\theta}{Y}$ of some parameter $\theta$ with standard deviation $\sigma_{\widehat{\theta}}=\sqrt{\VAR(\Est{\theta}{Y})}$, let us propose the test statistic 
\begin{equation}\label{eq-stat1}
\Ecart{\theta} :=\frac{\Est{\theta}{Y} - \theta}{\sigma_{\widehat{\theta}}} \mbox{ or }\frac{\Est{\theta}{Y} - \theta}{\widetilde{\sigma}_{\widehat{\theta}}}
\end{equation}
in the case where $\sigma_{\widehat{\theta}}$ or possibly some known asymptotic equivalent $\widetilde{\sigma}_{\widehat{\theta}}$ of $\sigma_{\widehat{\theta}}$ only depends on $\theta$, or
\begin{equation}\label{eq-stat2}
\EcartEst{\theta} :=\frac{\Est{\theta}{Y} - \theta}{\Est{\sigma_{\widehat{\theta}}}{Y}}
\end{equation}
otherwise, where $\Est{\sigma_{\widehat{\theta}}}{Y}$ is a consistent estimate of  $\sigma_{\widehat{\theta}}$.

In the large sample framework, the parameters $\mu$, $\sigma^2$, $d_\mu$, $r_\mu$, $d_{\sigma^2}$ and $r_{\sigma^2}$ fall into the case of equation~(\ref{eq-stat2}) and the corresponding statistic approximately follows a $\mathcal{N}(0,1)$ distribution. In this same context, the proportion parameter falls into the case of equation~(\ref{eq-stat1}) with $\sigma_{\widehat{p}}=\sqrt{p(1-p)/n}$.

In the Gaussian framework, the parameters $\mu$ and $d_\mu$ leading to the one-sample and two-sample t-tests also fall into the case of equation~(\ref{eq-stat2}). Let us now concentrate on parameters $\sigma^2$ and $r_{\sigma^2}$ (corresponding to the chi-square variance test and the Fisher test). It is known that the variance of \Est{\sigma^2}{Y^G} and an asymptotic equivalent of the variance of \Est{r_{\sigma^2}}{Y^G} are of the form
$$
\sigma^2_{\widehat{\sigma^2}}= \frac1n\left( \dot{\mu}_4-\frac{n-3}{n-1} \dot{\mu}_2^2 \right)
\; \mbox{ and } \;
\widetilde{\sigma}^2_{\widehat{r_{\sigma^2}}} = \frac{1}{n^{(1)}} \frac{\muC{4}{1}-(\muC{2}{1})^2}{(\muC{2}{2})^2} +
 \frac{1}{n^{(2)}} r_{\sigma^2}^2 \frac{\muC{4}{2}-(\muC{2}{2})^2}{(\muC{2}{2})^2}
$$
where $\dot{\mu}_k=E\Big((Y-E(Y))^k\Big)$ is the $k-$th centered moment of $Y$ (with our notation $\dot{\mu}_2=\sigma^2$). For Gaussian variables, $\dot{\mu}_4=3( \dot{\mu}_2)^2$ which leads to 
$$
\sigma^2_{\widehat{\sigma^2}} = \frac{2}{n-1} \sigma^4
\; \mbox{ and } \;
 \widetilde{\sigma}^2_{\widehat{r_{\sigma^2}}} =2 r_{\sigma^2}^2\left(\frac{1}{n^{(1)}}+ \frac{1}{n^{(2)}}\right).
$$
Now, in order to build a test, let us give the distributions of $\EcartG[Y^G]{\sigma^2}$ and $\EcartG[Y^G]{ r_{\sigma^2}}$ expressed in terms of $\Vn(\Vect{Y^G})$ and \Fn(\Vect{Y^G}):
$$
\EcartG[Y^G]{\sigma^2}:=\frac{ \Est{\sigma^2}{Y^G}-\sigma^2}{\sigma_{\widehat{\sigma^2}} }= \frac{ \frac{ \Vn(\Vect{Y^G}) }{n-1} \sigma^2-\sigma^2}{\sigma^2\sqrt{2/(n-1)}} \leadsto \frac{ \chi^2(n-1)-(n-1)}{\sqrt{2(n-1)}}
$$
and 
$$
\EcartG[Y^G]{r_{\sigma^2}}:=\frac{  \Est{r_{\sigma^2}}{Y^G}   -r_{\sigma^2}}{\widetilde{\sigma}_{\widehat{r_{\sigma^2}}} }=\frac{ r_{\sigma^2}\Fn(\Vect{Y^G})   -r_{\sigma^2}}{r_{\sigma^2} \sqrt{2/n^{(1)}+2/n^{(2)}}}  \leadsto  \frac{ \mathcal{F}(n^{(1)}-1,n^{(2)}-1) -1}{ \sqrt{2/n^{(1)}+2/n^{(2)}}}.
$$
One may propose a name for the previous two free distributions: centered reduced chi-square distribution and centered reduced Fisher distribution respectively denoted by $\chi^2_{cr}(\cdot)$ and $\mathcal{F}_{cr}(\cdot,\cdot)$. If these distributions were implemented such that one may evaluate quantiles and p-values, one could build two new tests directly based on \EcartG[Y^G]{{\sigma^2}} and \EcartG[Y^G]{r_{\sigma^2}}. Of course, these two new tests would be strictly equivalent to the classical chi-square variance test and Fisher test and would then fall into the same formalism.

Let us now comment on the concept of robustness in the Gaussian framework. In the particular case of mean hypothesis testing, this robustness is expressed by  the fact that $\EcartG[Y^G]{\mu}$ is equal to $\Ecart[Y^G]{\mu}$ with the same asymptotic distribution $\mathcal{N}(0,1)$. When considering the two new cases $\theta=\sigma^2$ or $\theta=r_{\sigma^2}$, $\EcartG[Y^G]{\theta}$ is no longer equal to $\Ecart[Y^G]{\theta}$ but have at least the same asymptotic distribution $\mathcal{N}(0,1)$. However, one can prove that $\EcartG[Y^G]{\theta}$ and $\Ecart[Y^G]{\theta}$ are asymptotically equivalent in probability, which may be viewed as some kind of robustness.

\section{Proofs} \label{sec-proofs}

In this section, we only prove (\ref{eq-tcl}). The results~(\ref{conv-Vn}) and~(\ref{conv-Fn}) are direct consequences. Recall that for each parameter, some assumptions are needed essentially in order to apply the central limit theorem. They are summarized in Tab.~\ref{tab-resTCL}.

\textbf{Parameter $\mu$:}\\

This is a direct application of the central limit theorem.

\textbf{Parameter $\sigma^2$:}

By definition
$$
\Est{\sigma^2}{Y} - \sigma^2 = \frac1{n-1} \sum_{i=1}^{n}\left( \left(Y_i-\overline{Y}\right)^2 -\sigma^2\right) =  \frac1{n-1} \sum_{i=1}^{n}\left( \left(Y_i-\mu \right)^2 -\sigma^2\right) - \frac{n}{n-1}\left( \overline{Y}-\mu\right)^2.$$
From the CLT, the law of large numbers and Slutsky's Theorem (see {\it e.g.} \cite{Ferguson96}), it comes that as $n\to +\infty$ 
$$
\sqrt{n} \left( \overline{Y}-\mu\right)^2 \stackrel{\mathbb{P}}{\longrightarrow} 0.
$$
Therefore, as $n\to+\infty$,
$$
\sqrt{n}\left( \Est{\sigma^2}{Y} - \sigma^2 \right)  \stackrel{d}{\longrightarrow} \mathcal{N}(0,\VAR((Y-\mu)^2)).
$$
Since $\sigma_{\widehat{\sigma^2}}:=\sqrt{\frac{\VAR((Y-\mu)^2)}{n}}$ can be consistently estimated by $\Est{{\sigma}_{\widehat{\sigma^2}}}{\Vect{Y}}:=\sqrt{\frac{\Est{\sigma^2_{\ddot{Y}}}{\ddot{\Vect{Y}}}}{n}}$, see Tab.~\ref{tab-resTCL}, we obtain as $n\to+\infty$
$$
\EcartEst{\sigma^2}:=\frac{\Est{\sigma^2}{Y} - \sigma^2}{ \Est{{\sigma}_{\widehat{\sigma^2}}}{\Vect{Y}} } \stackrel{d}{\longrightarrow} \mathcal{N}(0,1).
$$

\textbf{Parameter $d_\mu$:}\\

Recall that $n=n^{(1)}$ and $n^{(2)}=\alpha n^{(1)}$. As $n\to +\infty$,
\begin{eqnarray*}
\sqrt{n} \left( \Est{d_\mu}{Y}-d_\mu\right) &=& \sqrt{n} \left(\Est{\mu^{(1)}}{Y^{(1)}} - \mu^{(1)} \right) - \rho\times\frac{\sqrt{n\alpha}}{\sqrt{\alpha}} \left(\Est{\mu^{(2)}}{Y^{(2)}} - \mu^{(2)} \right) \\
&\stackrel{d}{\longrightarrow}& \mathcal{N}\left(0,\sigma^2_{(1)} + \rho^2\times \frac{\sigma^2_{(2)}}\alpha \right).
\end{eqnarray*}
Since $\sigma_{\widehat{d_\mu}}:=\sqrt{\frac{\sigma^2_{(1)} + \rho^2\times\frac{\sigma^2_{(2)}}\alpha}n}$ can be consistently estimated by 
$$\Est{\sigma_{\widehat{d_\mu}}}{Y}:=\sqrt{ \frac{\Est{\sigma_{(1)}^2}{Y^{(1)}}}{n^{(1)}} + \rho^2\times \frac{\Est{\sigma_{(2)}^2}{Y^{(2)}}}{n^{(2)}} },$$
we obtain as $n\to+\infty$
$$\EcartEst{d_\mu}:=\frac{\Est{d_\mu}{Y}-d_\mu}{\Est{\sigma_{\widehat{d_\mu}}}{Y}}\stackrel{d}{\longrightarrow} \mathcal{N}(0,1).$$

\textbf{Parameter $d_{\sigma^2}$:}\\

As $n \to +\infty$,
\begin{eqnarray*}
\sqrt{n} \left( \Est{d_{\sigma^2}}{Y}-d_{\sigma^2}\right) &=& \sqrt{n} \left(\Est{\sigma^2_{(1)}{(1)}}{Y^{(1)}} - \sigma^2_{(1)} \right) -\rho\times \frac{\sqrt{n\alpha}}{\sqrt{\alpha}} \left(\Est{\sigma^2_{(2)}}{Y^{(2)}} - \sigma^2_{(2)} \right) \\
&\stackrel{d}{\longrightarrow}& \mathcal{N}\left(0,\sigma^2_{\ddot{Y^{(1)}}} + \rho^2\times\frac{\sigma^2_{ \ddot{Y^{(2)}} }}\alpha 
\right).
\end{eqnarray*}
Since $\sigma_{\widehat{d_{\sigma^2}}}:=\sqrt{\frac{\sigma^2_{\ddot{Y^{(1)}}} + \rho^2\frac{\sigma^2_{ \ddot{Y^{(2)}} }}\alpha}n}$ can be consistently estimated by $$\Est{\sigma_{\widehat{d_{\sigma^2}}}}{Y} :=\sqrt{ \frac{\Est{\sigma^2_{\ddot{Y^{(1)}}}}{\ddot{Y^{(1)}}}}{n} + 
\rho^2\times\frac{\Est{\sigma^2_{\ddot{Y^{(2)}}}}{\ddot{Y^{(2)}}}}{n\alpha} }=
\sqrt{ \frac{\Est{\sigma^2_{\ddot{Y^{(1)}}}}{\ddot{Y^{(1)}}}}{n^{(1)}} + 
\frac{\Est{\sigma^2_{\ddot{Y^{(2)}}}}{\ddot{Y^{(2)}}}}{n^{(2)}} } ,$$
we obtain as $n\to +\infty$
$$
\frac{\Est{d_{\sigma^2}}{Y}-d_{\sigma^2}}{ \Est{\sigma_{\widehat{d_{\sigma^2}}}}{Y} } \stackrel{d}{\longrightarrow} \mathcal{N}(0,1).
$$

\textbf{Parameter $r_\mu$:} \\

Using Slutsky's Theorem, one may assert that as $n\to +\infty$
\begin{eqnarray*}
\sqrt{n} \left( \Est{r_\mu}{Y} -r_\mu\right) &=& 
\sqrt{n} \left( \frac{\Est{\mu^{(1)}}{Y^{(1)}}}{ \Est{\mu^{(2)}}{Y^{(2)}}} -   \frac{\mu^{(1)}}{\mu^{(2)}}\right) \\
&=& 
\sqrt{n} \left( \frac{ \Est{\mu^{(1)}}{Y^{(1)}}-\mu^{(1)}}{\Est{\mu^{(2)}}{Y^{(2)}}}
+\mu^{(1)} \left( 
\frac1{\Est{\mu^{(2)}}{Y^{(2)}}} - \frac1{\mu^{(2)}}
\right) \right) \\
&\stackrel{n\to+\infty}{\sim}& \sqrt{n} \left( \frac{ \Est{\mu^{(1)}}{Y^{(1)}}-\mu^{(1)}}{\mu^{(2)}}
+ \frac{\sqrt{n\alpha}}{\sqrt{\alpha}} \frac{\mu^{(1)}}{(\mu^{(2)})^2} \left( \mu^{(2)} - \Est{\mu^{(2)}}{Y^{(2)}}\right)
\right) \\
&\stackrel{d}{\longrightarrow} & \mathcal{N}\left( 0,
\frac{\sigma^2_{(1)}}{(\mu^{(2)})^2}+ \left( \frac{\mu^{(1)}}{(\mu^{(2)})^2}\right)^2 \frac{\sigma^2_{(2)}}\alpha\right).
\end{eqnarray*}

Since, $\sigma_{\widehat{r_\mu}}:=\sqrt{ \frac{\frac{\sigma^2_{(1)}}{(\mu^{(2)})^2}+ \left( \frac{\mu^{(1)}}{(\mu^{(2)})^2}\right)^2 \frac{\sigma^2_{(2)}}\alpha}n}$ can be consistently estimated by
\begin{eqnarray*}
\Est{\sigma_{\widehat{r_\mu}}}{Y}&:=&
\sqrt{ \frac{ \frac{\Est{\sigma^2_{(1)}}{Y^{(1)}} }{\Est{\mu^{(2)}}{Y^{(2)}}}
}{n} + \frac{ \left( \frac{\Est{\mu^{(1)}}{Y^{(1)}}}{\Est{\mu^{(2)}}{Y^{(2)}}^2}\right)^2  \Est{\sigma^2_{(2)}}{Y^{(2)}}
}{n\alpha}
}\\ &=& \frac1{\Est{\mu^{(2)}}{Y^{(2)}}} \sqrt{
\frac{\Est{\sigma^2_{(1)}}{Y^{(1)}}}{n^{(1)}} +
\Est{r_\mu}{Y} \frac{\Est{\sigma^2_{(2)}}{Y^{(2)}}}{n^{(2)}}
}\\
\end{eqnarray*}
we obtain as $n \to +\infty$,
$$
\frac{\Est{r_\mu}{Y} -r_\mu}{\Est{\sigma_{\widehat{r_\mu}}}{Y}} \stackrel{d}{\longrightarrow}  \mathcal{N}( 0,1).
$$

\textbf{Parameter $r_{\sigma^2}$:} \\

The proof follows the ideas developed for the parameters $\sigma^2$, $d_{\sigma^2}$ and $r_\mu$, and thus is left to the reader.

\section*{Acknowledgments}
We would like to thank Laurence Pierret for having kindly reread our English. She could in no way be held responsible for any remaining mistakes. 

\bibliographystyle{plainnat.bst}

\bibliography{asymp}

\end{document}